\documentclass[12pt]{elsarticle}
\usepackage{natbib}
\usepackage{amssymb}
\usepackage{amsmath}
\usepackage{graphicx}
\usepackage{bm}
\usepackage{epsfig}
\usepackage{multirow}
\usepackage{lineno}
\usepackage{array}
\usepackage{colortbl}
\usepackage{lscape}
\usepackage{tikz}
\usepackage{makecell}
\usepackage{subfigure}
\usepackage{float}
\usepackage{natbib}
\usepackage{lscape}
\usepackage{multirow}
\usepackage{array,multirow}
\usepackage{xcolor}
\usepackage{lipsum}
\usepackage{threeparttable}

\newcommand{\argmin}[1]{\underset{#1}{\operatorname{argmin}}}
\biboptions{sort&compress}
\begin{document}
\begin{frontmatter}
\title{Tests  for  Comparing Weighted Histograms. Review and Improvements}
\author{Nikolay D.~Gagunashvili\corref{cor1}}
\ead{nikolay@hi.is}
\cortext[cor1]{Tel.: +3545254000; fax: +3545521331}
\address{University of Iceland, S\ae mundargata 2, 101 Reykjavik, Iceland}
\begin{abstract}

Histograms with weighted entries are used to estimate probability density functions. Computer simulation is the main
application of this type of histograms. A review on chi-square  tests for comparing weighted histograms  is presented in this paper. Improvements to  these tests that have a size closer to its nominal value are proposed.  Numerical examples are presented for evaluation and demonstration of various applications of the tests.
\end{abstract}
\end{frontmatter}
\section{Introduction}
  A histogram with $m$ bins for a given probability density function (PDF)  $p(x)$ is used to estimate the probabilities $p_i$ that a random event  belongs to bin $i$:
\begin{equation}
p_i=\int_{S_i}p(x)dx, \; i=1,\ldots ,m. \label{p1}
\end{equation}
 Integration in (\ref{p1}) is carried out over  the bin  $S_i$ and $\sum_1^m p_i=1$.
A histogram can be obtained as a result of a random experiment with the PDF $p(x)$.

A frequently used technique in data analysis is to compare two distributions through  comparison of histograms.
The  hypothesis of homogeneity  states that two histograms
  represent random  values with  identical distributions \cite{cramer}.
  It is equivalent to the existing $m$ constants $p_1,...,p_m$,
 such that $\sum_{i=1}^{m} p_i=1$,
 and the probability  of  belonging  to the $i$$^{\textrm{th}}$  bin for some  measured value
 in both experiments is  equal to $p_i$.

Let us denote the numbers of random events belonging to the $i$$^{\textrm{th}}$  bin of the first and second histogram as $n_{1i}$ and $n_{2i}$, respectively. The total number of events in the histograms is equal to $n_j=\sum_{i=1}^{m}{n_{ji}}$, where $j = 1, 2$.

It has been shown by Pearson \cite{pearson} that the goodness of fit test  statistic
\begin{equation}
\sum_{i=1}^{m} \frac{(n_{ji}-n_jp_{i})^2}{n_jp_{i}} \label{basic}
\end{equation}
 has  approximately a $\chi^2_{m-1}$ distribution.
For two statistically independent histograms with probabilities $p_1,...,p_m$,  the statistic
\begin{equation}
\sum_{j=1}^2  \sum_{i=1}^{m} \frac{(n_{ji}-n_jp_{i})^2}{n_j p_{i}} \label{basic2}
\end{equation}
has approximately a $\chi^2_{2m-2}$ distribution.
 If the probabilities $p_1,...,p_m$ are not known, the estimation of $p_i$ is carried out by the following expression:
 \begin{equation}
 \hat{p}_i= \frac{n_{1i}+n_{2i}}{n_{1}+n_{2}} \, \text{ ,} \label{phat1}
 \end{equation}
 as shown in \cite{cramer}.
By substituting expression (\ref{phat1}) in (\ref{basic2}), the statistic
\begin{equation}
X^2=\sum_{j=1}^2  \sum_{i=1}^{m} \frac{(n_{ji}-n_j\hat{p}_{i})^2}{n_j \hat{p}_{i}}
=\frac{1}{n_1n_2} \sum_{i=1}^{m}{\frac{(n_2n_{1i}-n_1n_{2i})^2}{n_{1i}+n_{2i}}} \label{xsquar1}
\end{equation}
is obtained.
This statistic has approximately a $\chi^2_{m-1}$ distribution
 because  $m-1$ parameters  are  estimated \cite{cramer}.
 The statistic  (\ref {xsquar1}) was first  proposed in \cite{fisher} and is widely used to test the  hypothesis of homogeneity.

A weighted histogram or   a histogram with weighted events  \cite{gagu1, gagu2, gagu3} is used to estimate the probabilities $p_i$ (\ref{p1}) as well. 
The sum of weights of events for the bin $i$ is defined as:
 \begin{equation}
W_i=  \sum_{k=1}^{n_i}w_i(k), \label{ffffweight}
\end{equation}
where $n_i$ is the number of events in the bin $i$ and $w_i(k)$ is the weight of the $k$$^{\textrm{th}}$ event in the $i$$^{\textrm{th}}$ bin.  The statistic
\begin{equation}
\hat{p_i}=W_i/n  \label{west}
\end{equation}
is used to estimate $p_i$, where $n=\sum_{i=1}^{m}{n_i}$ is a total number of events for the histogram with  $m$ bins. Weights of events are chosen in such a way that the estimate  (\ref {west}) is unbiased:
\begin{equation}
 \textrm E [\hat{p_i}]=p_i.
 \end{equation}

 Because of the condition $\sum_ip_i=1$, we will further call the above defined weights ``normalized" as opposed to the unnormalized weights $\check w_i(k)$ which are $\check w_i(k)=const\cdot w_i(k)$.

 Comparison of two weighted histograms  and comparison of weighted and unweighted histograms as well as fitting weights of simulated random events to an experimental histogram are all important parts of data analysis. Tests for comparing  weighted  histograms have been developed in \cite{gagucom1,gagucom2} while tests for Poisson weighted histograms have been proposed  in \cite{zex}.

This paper is organized as follows. In Section \ref{hom_test_norm} generalization of the chi-square homogeneity test
 is discussed and improvements for the test are proposed.  A test for histograms with unnormalized weights as well as improvements of that test are discussed in Section \ref{hom_test_unnorm}. Tests for comparison  of two Poisson weighted histograms are discussed in  Section \ref{hom_pois}.  Restrictions  for chi-square test application are discussed  in Section
 \ref{rest}. Applications and verification of the tests are demonstrated using numerical examples in Section \ref{examples}.
\section{Homogeneity test for comparison two histograms with normalized weights}\label{hom_test_norm}

Let us consider two  histograms with normalized weights, and the subindex $j$ will be used to differentiate them. A total
 sum of weights of events $W_{ji}$ in the $i$$^{\textrm{th}}$ bin of the $j$$^{\textrm{th}}$ histogram $j=1,2$;  $i=1,\ldots,m$ can
  be considered as a sum of random variables
 \begin{equation}
W_{ji}= \sum_{k=1}^{n_{ji}}w_{ji}(k), \label{ffffweight}
\end{equation}
where the number of events  $n_{ji}$ is also a random value and the
weights $w_{ji}(k), k=1,...,n_{ji}$ are independent
 random variables with the same PDF for a given bin \cite{gagu1,gagu3}.
 Let us introduce a variable
\begin{equation}
r_{ji}=\textrm E \, [w_{ji}]/\textrm E[ \, w_{ji}^2], \label{rati}
\end{equation}
 which is the ratio of the first moment  to the second moment of the distribution of weights in the bin $i$. Let us estimate $r_{ji}$ using
\begin{equation}
\hat r_{ji}=\sum_{k=1}^{n_{ji}}w_{ji}(k)/\sum_{k=1}^{n_{ji}}w_{ji}^2(k).\label{rat4}
\end{equation}
As shown in \cite{gagu1} the statistic
\begin{equation}
 \frac{1}{n_j} \sum_{i \neq k} \frac{\hat r_{ji}W_{ji}^2}{p_{i}}+\frac{1}{n_j}
\frac{(n_j-\sum_{i \neq k}\hat r_{ji}W_{ji})^2}{1-\sum_{i \neq k}\hat r_{ji}
p_{i}}-n_j, \label{stddu}
\end{equation}
 where sums extend over all the bins $i$, except for the bin $k$, which
has  approximately  a $\chi^2_{m-1}$ distribution and is a generalization of the Pearson's statistic (\ref{basic})  \cite{robins,gagu1,gagu3}. It should be noted that it is only  valid  for the case when $1-\sum_{i \neq k}\hat r_{ji}p_{i}> 0$. The last inequality means that estimation of a covariance matrix for variables $W_{j1},...,W_{jk-1},W_{jk+1},...,W_{jm}$ is positive definite.

  The better power of test, as was shown in \cite{gagu3}, was achieved  for $k_j$, where
\begin{equation}
k_j = \argmin{i} \frac{\hat p_i}{\hat{r}_{ji}}. 
\end{equation}

\subsection{Median test statistic for comparison of weighted histograms with normalized weights}
Following \cite{gagucom1},
for two statistically independent histograms with probabilities $p_1,...,p_m$  the statistic has approximately a $\chi^2_{m-1}$ distribution:
\begin{equation}
\hat X_k^2=\sum_{j=1}^2\frac{1}{n_j} \sum_{i \neq k} \frac{\hat r_{ji}W_{ji}^2}{\hat p_{i}}+\sum_{j=1}^2\frac{1}{n_j}
\frac{(n_j-\sum_{i \neq k}\hat r_{ji}W_{ji})^2}{1-\sum_{i \neq k}\hat r_{ji}
\hat p_{i}}-\sum_{j=1}^2n_j .\label{stma}
\end{equation}
The probabilities $p_i$ are not known and estimators $\hat p_1,\ldots,\hat p_{k-1},\hat p_{k+1},\ldots,\hat p_m$ can be determined by minimizing (\ref{stma}) under the following constraints:
 \begin{equation}
  \hat p_i > 0,\,\,  1-\sum_{i \neq k}\hat p_i > 0,\,\, 1-\sum_{i \neq k}\hat r_{1i}\hat p_i > 0, \text { and } 1-\sum_{i \neq k}\hat r_{2i} \hat p_i>0\label{constr}.
 \end{equation}

   The problem to determine the estimators of the probabilities $\hat p_i$ by minimizing  (\ref{stma})  has been solved numerically by coordinate-wise optimization in \cite{gagucom1, gagucom2}.  For every step, the minimum for one probability with others fixed ones  can be found using the Brent algorithm \cite{brent}.

  A test statistic obtained as a median value of the formula
    (\ref{stma}) for a different choice of the excluded bin
\begin{equation}
\hat X^2_{Med}= \textrm {Med }\, \{\hat X_1^2,  \hat X_2^2,  \ldots , \hat
X_m^2\}\label{stdavu}
\end{equation}
was proposed in \cite{gagucom1,gagucom2} and has  approximately a $\chi^2_{m-1}$ distribution if the hypothesis of homogeneity is valid.

The  median is calculated for the set of  statistically dependent random variables $\hat X_i^2 $, with each variable  having  approximately  $\chi^2_{m-1}$  distribution \cite{robins, gagu3}. The median statistic (\ref{stdavu}) coincides with the statistic (\ref{xsquar1}) in case of  two histograms with unweighted entries.
Numerical investigations of the median tests (see   Section 6.1 and  Ref. \cite{gagu1})   show that  the size of the test (\ref{stdavu}) exceeds slightly its nominal value   making it  the main disadvantage of this approach. The question, what deviation from the nominal size is acceptable for  chi-square methods, has different answers.
 
 In the classical work dedicated to chi-square tests  \cite{cochran}   disturbance is regarded as unimportant when the nominal size of a test is $5\%$,  with  the exact size lying between $4\%$ and $6\%$, and when the nominal size of a test is $1\%$, with the exact size   lying  between  $0.7\%$ and $1.5\%$. According to  this criteria the disturbance of the median test can be considered  unimportant.

However,  according to   \cite{zex},  the disturbance of the median statistics is important.   The authors of  \cite{zex}  have proposed tests for comparison of   histograms of an {\it equivalent number of  unweighted events}   with false interpretation of these  tests as tests for  histograms with weighted entries.  The methods  from \cite{zex} are discussed  in section 4 with numeric evaluation shown in subsection 6.2.1.
\subsection{New test statistic for comparison of weighted histograms with normalized weights}

The median test (\ref{stdavu}) can be improved by using the results for goodness of fit test for weighted histograms \cite{gagu3}.

The new test statistic is

\begin{equation}
\hat X^2=\sum_{j=1}^2\frac{1}{n_j} \sum_{i \neq k_j} \frac{\hat r_{ji}W_{ji}^2}{\hat p_{i}}+\sum_{j=1}^2\frac{1}{n_j}
\frac{(n_j-\sum_{i \neq k_j}\hat r_{ji}W_{ji})^2}{1-\sum_{i \neq k_j}\hat r_{ji}
\hat p_{i}}-\sum_{j=1}^2n_j .\label{stdd3}
\end{equation}
The estimation of the probabilities 
$\hat p_1,\ldots,\hat p_m$ is determined by minimizing (\ref{stdd3}) under the following constraints:
 \begin{equation}
  \hat p_i > 0,\,\,  \sum_{i}\hat p_i = 1,\,\, 1-\sum_{i \neq k_1}\hat r_{1i}\hat p_i > 0, \text { and } 1-\sum_{i \neq k_2}\hat r_{2i} \hat p_i>0\label{constr},
 \end{equation}
where $k_j$ is defined as
\begin{equation}
k_j = \argmin{i} \frac{\hat p_i}{\hat{r}_{ji}}.
\end{equation}
The test statistic asymptotically has a $\chi^2_{m-1}$ distribution
 and a size closer to its nominal value than the test  (\ref{stdavu}) if the hypothesis of homogeneity is valid. 

The bin $k_j$ with the lowest information content  is excluded to get the robust statistic  $\hat X^2$ and it  is plausible that the test (\ref{stdd3}) has higher power than the median test (\ref{stdavu}). Detail explanation of this choice is  presented in Subsection 2.3 of \cite{gagu3}.
\section{Homogeneity  test for histograms with unnormalized weights}\label{hom_test_unnorm}

In practice one is often confronted with  cases when a histogram is defined up to an unknown normalization constant. Let us denote  bin
content of histograms with unnormalized weights as $\check{W}_{ji}$, then
$W_{ji}=\check{W}_{ji}C_j$, and the test statistic (\ref{stddu}) can be
written as
\begin{equation}
 \frac{C_j}{n_j} \sum_{i \neq k} \frac{\check{r}_{ij}\check{W}_{ji}^2}{p_{i}}+\frac{1}{n_j}
\frac{(n_j-\sum_{i \neq k}\check{r}_{ji}\check{W}_{ji})^2}{1-C_j^{-1}\sum_{i
\neq k}\check{r}_{ji} p_{i}}-n_j,\label{stddc}
\end{equation}
with $\check{r}_{ji}=C_jr_{ji}$.
An estimator $\hat C_{jk}$  for the constant $C_j$ \ is
found  in \cite{gagu1} by minimizing (\ref{stddc}) and is equal to
\begin{equation}
\hat C_{jk}=\sum_{i \neq k}\check{r}_{ji}p_{i}+\sqrt{\frac{\sum_{i
\neq k}\check{r}_{ji}p_{i}}{\sum_{i \neq
k}\check{r}_{ji}\check{W}_{ji}^2/p_{i}}}(n_j-\sum_{i \neq
k}\check{r}_{ji}\check{W}_{ji}). \label{const}
\end{equation}
Substituting (\ref{const}) for (\ref{stddc}) and replacing $\check{r}_{ji}$ with the estimate  $\hat{\check{r}}_{ji}$  we get the test statistic

\begin{equation}
 \frac{\hat C_{jk}}{n_j} \sum_{i \neq k} \frac{\hat{\check{r}}_{ji}\check{W}_{ji}^2}{p_{i}}+\frac{1}{n_j}
\frac{(n_j-\sum_{i \neq k}\hat{\check{r}}_{ji}\check{W}_{ji})^2}{1-C_{j}^{-1}\sum_{i
\neq k}\hat{\check{r}}_{ji} p_{i}}-n_j,\label{stddc4}
\end{equation}
The estimate $\hat{\check{r}}_{ji}$ in (\ref{stddc4}) is calculated in the same way as the estimate $\hat{r}_{ji}$ in (\ref{rat4}).

The statistic (\ref{stddc4}) has approximately a $\chi^2_{m-2}$ distribution.


\subsection{Median test statistic for comparison of weighted histograms with unnormalized  weights}

Following \cite{gagucom1}, for two statistically independent histograms with probabilities $p_1,..., p_m$, the statistic
\begin{equation}
\hat {\check{X}}_k^2=\sum_{j=1}^2 \frac{\hat C_{jk}}{n_j} \sum_{i \neq k} \frac{\hat{\check{r}}_{ji}\check{W}_{ji}^2}{\hat p_{i}}+\frac{1}{n_j}
\frac{(n_j-\sum_{i \neq k}\hat{\check{r}}_{ji}\check{W}_{ji})^2}{1-\hat C_{jk}^{-1}\sum_{i
\neq k}\hat{\check{r}}_{ji}\hat p_{i}}-n_j,\label{st5}
\end{equation}
has approximately a $\chi^2_{m-2}$ distribution. An estimation of the probabilities 
$\hat p_1,\ldots,\hat p_{k-1},\hat p_{k+1},\ldots,\hat p_m$  can be found by minimizing (\ref{st5}) under the following constraints:
\begin{equation}
  \hat p_i > 0,\,\,  1-\sum_{i \neq k}\hat p_i > 0,\,\, 1-\hat C_{1k}^{-1}\sum_{i \neq k}\hat r_{1i}\hat p_i > 0, \text { and } 1-\hat C_{2k}^{-1}\sum_{i \neq k}\hat r_{2i} \hat p_i>0\label{constr}.
 \end{equation}
 The probabilities $\hat p_{i}$ can be calculated numerically in the same way as described in Section \ref{hom_test_norm}.
A test statistic that is ``invariant" to the choice of the excluded bin can be obtained again as a
 median value of (\ref{stdav2}) for all possible choices of the excluded bin
\begin{equation}
\hat {\check{X}}^2_{Med}= \textrm {Med }\, \{\hat {\check{X}}_1^2, \hat
{\check{X}}_2^2, \ldots , \hat {\check{X}}_m^2\}.\label{stdav2}
\end{equation}
The statistic $_1\hat {\check{X}}^2_{Med}$ for the case of comparing two histograms with normalized and unnormalized  weights can be given by the same formulas (\ref{st5}--\ref{stdav2})  with $C_{1k}\equiv1$.

Both statistics $\hat {\check{X}}^2_{Med}$ and  $_1\hat {\check{X}}^2_{Med}$ have approximately a $\chi^2_{m-2}$  distribution if the hypothesis of homogeneity is valid.


\subsection{New test statistic for comparison of weighted histograms with unnormalized  weights}
The median test (\ref{stdav2}) can be improved by using the results for goodness of fit test for weighted histograms \cite{gagu3}.

A new test statistic is
\begin{equation}
\hat {\check{X}}^2=\sum_{j=1}^2 \frac{\hat C_{jk_j}}{n_j} \sum_{i \neq k_j} \frac{\hat{\check{r}}_{ji}\check{W}_{ji}^2}{\hat p_{i}}+\frac{1}{n_j}
\frac{(n_j-\sum_{i \neq k_j}\hat{\check{r}}_{ji}\check{W}_{ji})^2}{1-\hat C_{jk_j}^{-1}\sum_{i
\neq k_j}\hat{\check{r}}_{ji}\hat p_{i}}-n_j.\label{stddc5}
\end{equation}
 Estimation of the probabilities $\hat p_{i}$ can be determined by minimizing  (\ref{stddc5}) under the following constraints:
\begin{equation}
  \hat p_i > 0,\,\,  \sum_{i}\hat p_i = 1,\,\, 1-\hat C_{1k_1}^{-1}\sum_{i \neq k_1}\hat r_{1i}\hat p_i > 0, \text { and } 1-\hat C_{2k_2}^{-1}\sum_{i \neq k_2}\hat r_{2i} \hat p_i>0\label{constr},
 \end{equation}
where $k_j$ is defined as
\begin{equation}
k_j = \argmin{i} \frac{\hat p_i}{\hat{r}_{ji}}. \label{cho}
\end{equation}

The test statistic asymptotically has a $\chi^2_{m-2}$ distribution
 and a size closer to its nominal value.  It is plausible that the test (\ref{stddc5}) has higher power than the test (\ref{stdav2}).

The statistic $_1\hat {\check{X}}^2$ for the case of comparing two  histograms with normalized and unnormalized
weights can be given by the same formulas (\ref{stddc5}--\ref{cho})  with $C_{1k_1}\equiv1$.

Both statistics $\hat {\check{X}}^2$ and  $_1\hat {\check{X}}^2$  have approximately a  $\chi^2_{m-2}$  distribution if the hypothesis of homogeneity is valid.

 \section{Test for comparison of  weighted Poisson histograms}\label{hom_pois}

A Poisson  histogram \cite{kendall,cousine} is defined as a histogram with  multi-Poisson distributions of a number of events for bins:
\begin{equation}
P(n_1,\ldots ,n_m)=\prod_{i=1}^m e^{-n_0p_i}(n_0p_i)^{n_i}/n_i!,\label{pooo}
\end{equation}
where $n_0$ is a free parameter.

 The probability distribution function (\ref{pooo}) can be represented as a product of two probability functions: a Poisson probability distribution function  for a number of events $n$  with the parameter  $n_0$ and a multinomial probability distribution  function of a number of events for bins of the  histogram, with a total number of events equal to $n$ \cite{kendall,cousine}:
\begin{equation}
P(n_1,\ldots ,n_m)=e^{-n_0}(n_0)^{n}/n!\times\frac{n!}{n_1!n_2! \ldots n_m!} \; p_1^{n_1}\ldots p_m^{n_m}.
\end{equation}

 A Poisson histogram can be obtained as a result of two random experiments, namely when a first experiment with a Poisson probability distribution  function gives us a total number of events in the histogram $n$   and then a histogram is obtained as a result of a random experiment with a PDF $p(x)$ and with a total number of events equal to $n$.

  The concept  of an  {\it equivalent number of unweighted events}   has been introduced  in  \cite{zex}. An {\it equivalent number of unweighted events} for $i$$^{\textrm{th}}$   bin of weighted histogram is    $W_{i}r_i$.  
The authors proposed two test statistics for comparison of  histograms with {\it equivalent number of unweighted events} contents of bins. These statistics were interpreted in \cite {zex} as statistics for comparison of original Poisson weighted histograms.

\subsection{First statistic for comparing Poisson weighted  histograms}
The first statistic $X^2_{p1}$,  in our notation, can be written as
\begin{equation}
X^2_{p1}=C^{-1}\sum_{i=1}^m\frac{( W_{1i}-C W_{2i})^2}{W_{1i}r_{2i}^{-1}+W_{2i}r_{1i}^{-1}}.\label{pois}
\end{equation}
 The parameter $C$ \cite{zex} is  taken equal to
\begin{equation}
C= \frac{\sum W_{1i}}{\sum W_{2i}}.
\end{equation}
The statistic (\ref{pois}) according to \cite{zex} has a $\chi^2_{m}$ distribution if the hypothesis of homogeneity is valid.\\
\subsection{Second statistic for comparing Poisson weighted  histograms }
The parameter $C$ can also  be estimated \cite{zex}.
Here an estimator $\hat C$   was found  by  minimizing (\ref{pois}) and  is equal to
\begin{equation}
\hat C= \sqrt{\sum \frac {W_{1i}^2}{W_{1i}r_{2i}^{-1}+W_{2i}r_{1i}^{-1}} \Big( \sum \frac {W_{2i}^2}{W_{1i}r_{2i}^{-1}+W_{2i}r_{1i}^{-1}}\Big)^{-1}}.
\end{equation}

The second statistic
\begin{equation}
X^2_{p2}=\hat C^{-1}\sum_{i=1}^m\frac{( W_{1i}-\hat C W_{2i})^2}{W_{1i}r_{2i}^{-1}+W_{2i}r_{1i}^{-1}}\label{pois2}
\end{equation}
has a $\chi^2_{m-1}$ distribution if the hypothesis of homogeneity is valid \cite{zex}.

\section{Restrictions of chi-square test applications}\label{rest}

 The use of the chi-square test $X^2$ (\ref{xsquar1}) for the histograms with unweighted entries   is inappropriate
 if any expected frequency $n_1\hat p_i$ or  $n_2\hat p_i<1$  or if the total number of bins with  the expected frequency   $n_1\hat p_i$ or $n_2\hat p_i<5$  exceeds $20\%$ of the total number $(2m)$  of bins \cite{moore,cochran}.

 Restrictions for weighted histograms can be obtained   by replacing the above mentioned  expected frequencies with expected frequencies of the {\it equivalent number of unweighted events}.
  For the test $\hat X^2$(\ref{stdd3}) they must be replaced  with  $n_1\hat p_i\hat r_{1i}$ and $n_2\hat p_i\hat r_{2i}$, while for the test $\hat {\check{X}}^2$ (\ref{stddc5})  with $n_1\hat p_i\hat r_{1i}/C_{1k_1}$ and  $n_2\hat p_i\hat r_{2i}/C_{2k_2}$.


\section{Evaluation of the tests' sizes and power}\label{examples}

The hypothesis  of homogeneity $H_0$ is rejected if the value of the test statistic $\hat X^2$ is above
 a given threshold. The threshold $k_{\alpha }$ for a given nominal
size of the test $\alpha$ can be  defined from the equation
\begin{equation}
\alpha = P\,(\chi^2_l>k_{\alpha})=\int_{k_{\alpha}}^{+\infty}
\frac{x^{l/2-1} e^{-x/2}}{2^{l/2}\Gamma(l/2)}dx, \label {kalfa}
\end{equation}
where $l=m-1$.

Let us define the test size $\alpha_s$ for a given nominal
size of the test $\alpha$ as the probability
\begin{equation}
\alpha_s =  P\,(\hat X^2>k_{\alpha}|H_0),\label {alfas}
\end{equation}
i.e.\ the probability that the hypothesis $H_0$ will be rejected if the distribution of the
weights $W_{ji}$,  $j=1,2;\,\, i=1,...,m$, for the bins of the histograms satisfies the
hypothesis $H_0$. The deviation of a  test size from its nominal value  is an important test characteristic.

A second important characteristic of the test is its power $\beta$
\begin{equation}
\beta = P\,(\hat X^2 >k_{\alpha}|H_a), \label {beta}
\end{equation}
i.e.\  the probability that the hypothesis of homogeneity $H_0$ will be rejected if the distributions
 of the weights $W_{ji}$, $j=1,2$; $i=1,...,m$ of the compared histograms do not satisfy the hypothesis $H_0$.

The same definitions with $l=m-2$ in the formula (\ref {kalfa}) can be used for the test statistic $\hat{\check{X}}^2$ (\ref{stddc5}).

Let us consider an example of  a weighted histogram for estimation of  the probability $p_i$
(\ref{p1}) for a given PDF  $p(x)$  in the form
\begin{equation}
p_i= \int_{S_i}p(x)dx = \int_{S_i}w(x)g(x)dx, \label{weightg}
\end{equation}
where
\begin{equation}
w(x)=p(x)/g(x) \label{fweightg}
\end{equation}
 is a weight function and $g(x)$ is some other PDF. The function $g(x)$ must be $>0$ for the points $x$, where $p(x)\neq 0$. The weight is equal to 0  if $p(x)=0$ \cite{sobol}.

 A weighted histogram is a histogram obtained from a random experiment with the PDF $g(x)$, and the weights of the events are calculated according to (\ref{fweightg}).

To evaluate a size and power of the tests let us take the distribution
\begin{equation}
p(x)\propto \frac{2}{(x-10)^2+1}+\frac{1}{(x-14)^2+1} \label{weight}
\end{equation}
 defined on the interval $\{4, 16\}$ and represented by two Breit-Wigner
  peaks \cite{breit}.
  Three cases of the PDF $g(x)$ can be
  considered (Fig. 1):
\begin{equation}
g_1(x)=p(x)   \label{prc}
\end{equation}
\begin{equation}
g_2(x)=1/12  \label{flat}
\end{equation}
\begin{equation}
g_3(x)\propto\frac{2}{(x-9)^2+1}+\frac{2}{(x-15)^2+1} \label{real}
\end{equation}

The distribution $g_1(x)$ (\ref{prc}) results in  a  histogram with unweighted entries, while the distribution $g_2(x)$ (\ref{flat}) is a uniform distribution
 on the interval $\{4, 16\}$. The distribution $g_3(x)$ (\ref{real}) has the same form of parametrization as $p(x)$ (\ref{weight}), but with different values of the parameters.

 Sizes of the tests for histograms with a number of bins equal to 5 and different weighted
 functions were calculated for the nominal size $\alpha$ equal to 0.05. 

Calculations of the test
  sizes $\alpha_s$   were carried out using the Monte Carlo method with 10\,000 runs, therefore it is reasonable to test the hypothesis $H_0^{(1)}:\alpha_s=0.05$ against the
  alternative $H_a^{(1)}: \alpha_s \neq 0.05$. For this purpose $z$ statistics can be used \cite{moore}
\begin{equation}
z=(\hat{\alpha}_s-0.05)/\sqrt{\frac{0.05\times(1-0.05)}{10\,000}},
\end{equation}
where $\hat{\alpha}_s$ is an estimated  value of $\alpha_s$.
If the null hypothesis is true then this test statistic has a standard normal distribution. 
 For the standard normal distribution, 2.5\% of the values lie below the critical value of $-1.959964$, and 2.5\% lie above 1.959964. Therefore, if a 2-sided hypothesis test is conducted with  a significance level  equal to 0.05,  $H_0^{(1)}$ is accepted when $|z| \leq 1.959964$ or  $0.045728\leq\hat{\alpha}_s\leq 0.054272$.  

  The results of   calculation for a pair of histograms with either normalized weights or unnormalized weights as
    well as for two histogram with normalized and unnormalized weights are presented in Tables~\ref{tab:result1}--\ref{tab:result3c} for different weight functions, and different total number of events.
To calculate  sizes of  tests two statistically independent weighted histograms were simulated. 
 The distribution $p(x)$  (\ref{weight}) was used for simulation of the  first weighted histogram and  the same distribution $p(x)$  for  simulation of the second one. 

  Weights  $p(x)/g_i(x)$  where  used for  histograms with normalized weighted entries  as well as  weights  $2p(x)/g_i(x)$  and  $3p(x)/g_j(x)$  for histograms with unnormalized weighted entries.

 Powers of the tests were investigated for slightly different values of the amplitude of the second  peak of the specified probability  distribution function (Fig. 2):
\begin{equation}
p_0(x)\propto \frac{2}{(x-10)^2+1}+\frac{1.15}{(x-14)^2+1}. \label{weight3}
\end{equation}
%
\subsection{Tests for histograms with a multinomial distribution of events} 
A size of the tests was calculated for a different total
number of events $n_1$ and $n_2$  in five bin histograms. In the following, numerical examples demonstrate applications of:
\begin{itemize}
  \item The median test $X^2_{Med}$(\ref{stdavu}) and  the new  test  $\hat X^2$(\ref{stdd3}) for comparison of weighted histograms with normalized
weights (Table~\ref{tab:result1});
  \item  The median test $\hat {\check{X}}^2_{Med}$ (\ref{stdav2}) and the new test $\hat {\check{X}}^2$(\ref{stddc5}) for comparison of weighted histograms with unnormalized weights (Table~\ref{tab:result2});
  \item  The median test   $_1\hat {\check{X}}^2_{Med}$ (\ref{stdav2})  and the new test  $_1\hat {\check{X}}^2$(\ref{stddc5}) for comparison of a weighted histogram with normalized weights and a histogram with unnormalized weights (Table~\ref{tab:result3}).
\end{itemize}
\vspace {-1.5cm}
\begin{figure}[H]
\centering \vspace*{-0.9 cm} \hspace*{-0.7 cm}
\includegraphics[width=1.15 \textwidth]{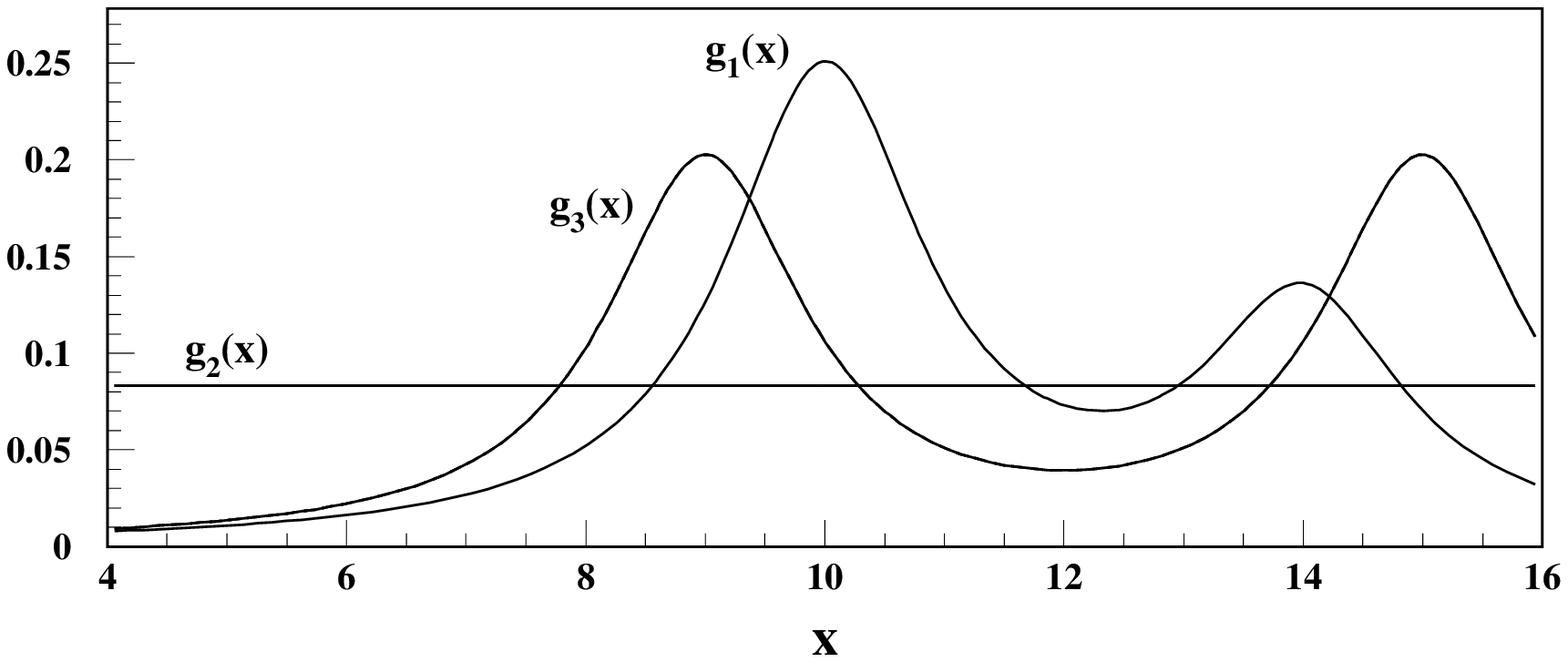}\\
\vspace*{-8. cm}
  \caption {Probability density functions $g_1(x)=p(x)$, $g_2(x)=1/12$ and $g_3(x)$.}
\vspace*{1 cm }
\centering \vspace*{-2.cm}
\hspace*{-0.7 cm}
  \includegraphics[width=1.15 \textwidth]{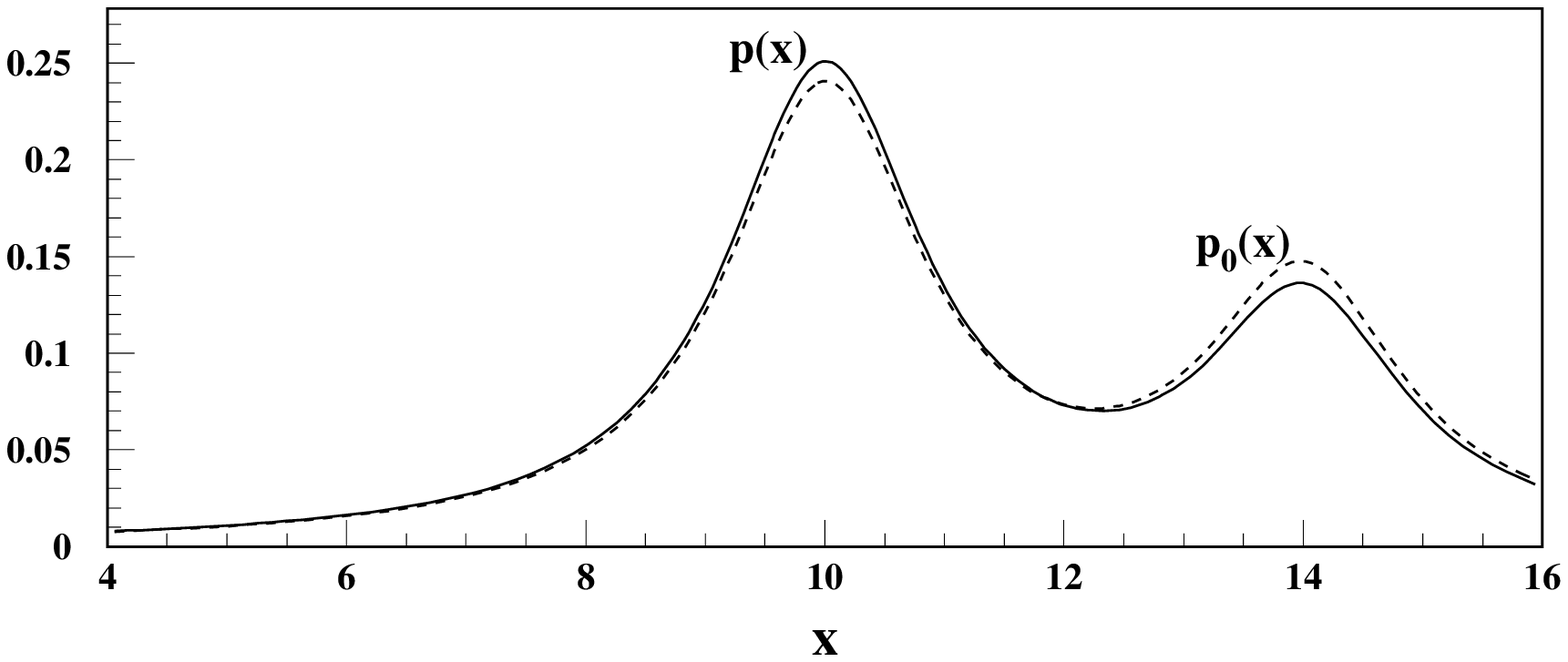}\\
\vspace*{-8. cm}
\caption {Probability density functions $p(x)$
(solid line)  and $p_0(x)$ (dashed line).} \vspace*{3 cm }
\end{figure}
\vspace*{-4.4 cm}
 \begin{table}[H]
\centering
\caption{ Sizes $\hat\alpha_s$  of the  test  $\hat X^2_{Med}$ (\ref{stdavu}) for comparison of two histograms with normalized weighted entries (left panel) and sizes of the new test $\hat X^2$ (\ref{stdd3})  (right panel) for different pairs of weights (last column) and  numbers of events $n_1$, $n_2$. Sizes of the  tests that do not satisfy the hypothesis $\alpha_s=5\%$ with a significance level equal to $0.05$ ($\hat\alpha_s>5.4\%$ or $\hat\alpha_s<4.6\%$) are highlighted with gray.}
\label{tab:result1}
\vspace *{0.2 cm}
\tiny
\begin{tabular}{r|rrrrrr||rrrrrr|c}
& \multicolumn{6}{c||}{$n_2$} & \multicolumn{6}{c|}{$n_2$} & \\
$n_1$&  200 &  400 &  800 & 1600 & 3200 & 6400 &  200 &  400 &  800 & 1600 & 3200 & 6400&$w(x)$\\
\hline
   200 &   4.9 & \cellcolor[gray]{.8}  4.5 &   4.7 &   4.6 &   5.2 &   4.9 & \cellcolor[gray]{.8} 4.2 &  4.7 &  4.6 &  4.9 & 4.6 &  4.8&  \\
   400 &   4.7 &   4.7 &   4.8 &   4.7 &   5.1 &   5.0 & \cellcolor[gray]{.8} 4.4 &  4.7 & \cellcolor[gray]{.8} 4.4 &  4.8 &  5.2 &  5.2& {$1$}\\
   800 &   4.9 &  \cellcolor[gray]{.8} 4.5 &   4.7 &   4.9 &   5.3 & \cellcolor[gray]{.8}  4.4 &  4.8 &  4.7 & \cellcolor[gray]{.8} 4.5 &  5.0 &  5.3 &  4.9&\multirow{2}{*} {$\&$} \\
  1600 &   5.2 &   4.7 &   5.3 &   4.7 & \cellcolor[gray]{.8}  5.5 &   5.1 &  \cellcolor[gray]{.8}4.4 &  4.9 &  5.0 &  4.9 &  4.9 &  5.0&\\
  3200 &   4.9 &   5.0 &   5.0 &   4.7 &   4.8 &   4.9 &  5.2 &  5.3 &  5.0 &  4.9 &  4.9 &  4.8& $1$\\
  6400 &   5.2 &   5.3 &   5.0 &   5.0 &   5.2 &   5.3 &  5.2 &  5.1 &  4.8 & \cellcolor[gray]{.8} 5.5 &  4.8 &  5.0&\\
\hline
   200 &  \cellcolor[gray]{.8} 5.5 & \cellcolor[gray]{.8}  5.7 &   5.1 &   5.0 &   5.1 &   5.1 &  4.7 &  4.9 &  5.2 &  5.0 &  5.0 &  4.6&\multirow{3}{*}{$1$} \\
   400 & \cellcolor[gray]{.8}  5.5 &  \cellcolor[gray]{.8} 5.7 &   5.4 &  \cellcolor[gray]{.8} 5.8 &   5.2 &   5.2 &  5.2 &  5.4 &  5.0 &  5.1 &  4.8 &  5.0& \\
   800 & \cellcolor[gray]{.8}  5.7 &   5.8 &  \cellcolor[gray]{.8} 5.5 &  \cellcolor[gray]{.8}  5.8 &   4.8 &   5.0 &  5.1 &  5.1 &  5.1 &  5.1 &  5.2 &  5.2&\multirow{2}{*}{$\&$} \\
  1600 &   5.3 & \cellcolor[gray]{.8}  5.5 & \cellcolor[gray]{.8}  5.7 &   5.2 &   4.9 &   4.9 &  5.4 & \cellcolor[gray]{.8} 5.5 &  5.4 &  5.1 &  5.2 &  5.2& \multirow{3}{*}{$\frac{p(x)}{g_2(x)}$}\\
  3200 &   5.4 &   5.3 &   5.4 & \cellcolor[gray]{.8}  5.6 &   5.3 &   5.4 & \cellcolor[gray]{.8} 5.6 & \cellcolor[gray]{.8} 5.5 &  5.1 &  4.9 &  4.8 &  4.8& \\
  6400 &  \cellcolor[gray]{.8} 5.6 & \cellcolor[gray]{.8}  5.5 &   5.4 &   5.3 &   5.3 &   5.3 & \cellcolor[gray]{.8} 5.5 &  4.8 &  5.1 &  5.1 &  5.1 &  5.0&\\
\hline
   200 &   5.0 &   4.8 &   5.0 &   4.9 &   4.8 &   5.3 &  5.3 &  5.0 &  4.7 &  4.7 &  4.7 &  4.7& \multirow{3}{*}{$1$}\\
   400 &   5.3 &   5.1 &   5.0 &   5.2 &   5.4 &   5.1 & \cellcolor[gray]{.8} 5.7 & \cellcolor[gray]{.8} 5.6 &  5.4 &  5.1 &  4.9 &  4.9& \\
   800 &   5.4 &  \cellcolor[gray]{.8} 5.6 &   5.2 &  \cellcolor[gray]{.8} 5.5 &   5.2 &   5.0 & \cellcolor[gray]{.8} 5.5 &  4.8 &  5.0 &  5.0 &  4.9 & \cellcolor[gray]{.8} 4.5&\multirow{2}{*}{$\&$}\\
  1600 & \cellcolor[gray]{.8}  5.7 &   5.4 &  \cellcolor[gray]{.8} 5.5 &   5.3 &  \cellcolor[gray]{.8} 5.6 & \cellcolor[gray]{.8}  5.5 & \cellcolor[gray]{.8} 5.7 & \cellcolor[gray]{.8} 5.5 &  5.1 &  5.0 &  5.3 &  5.2&\multirow{3}{*}{$\frac{p(x)}{g_3(x)}$}\\
  3200 & \cellcolor[gray]{.8}  5.9 &   5.3 &  \cellcolor[gray]{.8} 5.6 &  \cellcolor[gray]{.8} 5.5 &   5.3 &   5.3 & \cellcolor[gray]{.8} 6.2 & \cellcolor[gray]{.8} 5.5 &  5.4 &  5.1 &  5.1 &  5.1&\\
  6400 & \cellcolor[gray]{.8}  6.1 & \cellcolor[gray]{.8}  5.6 &   5.4 &   5.4 &  \cellcolor[gray]{.8} 5.5 &   5.3 & \cellcolor[gray]{.8} 5.6 &  \cellcolor[gray]{.8}6.2 &  5.1 &  5.0 &  5.0 &  5.2&\\
\hline
   200 &  \cellcolor[gray]{.8} 5.5 &   5.2 &  \cellcolor[gray]{.8} 5.5 & \cellcolor[gray]{.8}  6.1 & \cellcolor[gray]{.8}  5.9 & \cellcolor[gray]{.8}  5.8 & \cellcolor[gray]{.8} 5.7 &  5.3 & \cellcolor[gray]{.8} 5.5 &  5.2 & \cellcolor[gray]{.8} 5.5 &  5.0& \multirow{3}{*}{$\frac{p(x)}{g_2(x)}$}\\
   400 &   5.3 &   5.3 &  \cellcolor[gray]{.8} 5.8 & \cellcolor[gray]{.8}  5.5 & \cellcolor[gray]{.8}  5.7 &   5.3 & \cellcolor[gray]{.8} 5.8 &  5.1 &  5.0 &  5.1 &  4.7 &  5.0&\\
   800 & \cellcolor[gray]{.8}  6.0 &  \cellcolor[gray]{.8} 6.1 &  \cellcolor[gray]{.8} 5.5 &   5.3 &   5.2 &  \cellcolor[gray]{.8} 5.8 & \cellcolor[gray]{.8} 5.9 &  5.4 &  5.0 &  5.2 &  5.2 &  5.4&\multirow{2}{*}{$\&$}\\
  1600 &  \cellcolor[gray]{.8} 5.7 &  \cellcolor[gray]{.8} 5.5 &  \cellcolor[gray]{.8} 5.7 &   5.1 &   5.2 &   5.2 &  5.3 &\cellcolor[gray]{.8}  5.6 &  5.2 &  4.8 &  5.2 &  4.7& \multirow{3}{*}{$\frac{p(x)}{g_2(x)}$}\\
  3200 & \cellcolor[gray]{.8}  5.7 &   5.3 &   5.1 &  \cellcolor[gray]{.8} 5.5 &   5.1 &  \cellcolor[gray]{.8} 5.6 & \cellcolor[gray]{.8} 5.5 &  5.3 &  4.7 &  5.1 &  4.6 & \cellcolor[gray]{.8} 4.4&\\
  6400 &  \cellcolor[gray]{.8} 5.5 & \cellcolor[gray]{.8}  5.6 & \cellcolor[gray]{.8}  5.8 & \cellcolor[gray]{.8}  5.6 &   5.3 &   5.0 &  5.4 &  5.2 &  5.0 &  4.9 &  5.1 &  5.2&\\
\hline
   200 &  \cellcolor[gray]{.8} 5.7 &  \cellcolor[gray]{.8} 5.8 & \cellcolor[gray]{.8}  6.1 & \cellcolor[gray]{.8}  5.7 & \cellcolor[gray]{.8}  5.6 &   \cellcolor[gray]{.8}5.6 & \cellcolor[gray]{.8} 5.7 & \cellcolor[gray]{.8} 5.7 & \cellcolor[gray]{.8} 5.6 &  5.4 &  5.1 &  5.4&\multirow{3}{*}{$\frac{p(x)}{g_2(x)}$}\\
   400 & \cellcolor[gray]{.8}  5.5 &   5.0 &   5.1 & \cellcolor[gray]{.8}  5.6 &   5.2 & \cellcolor[gray]{.8}  5.8&  5.2 &\cellcolor[gray]{.8}  5.6 & \cellcolor[gray]{.8} 5.5 &  5.1 &  5.2 &  4.7& \\
   800 &   5.4 &  \cellcolor[gray]{.8} 5.6 &   5.2 & \cellcolor[gray]{.8}  5.6 & \cellcolor[gray]{.8}  5.5 &  \cellcolor[gray]{.8} 5.5 & \cellcolor[gray]{.8} 5.9 &  5.0 &  5.2 &  4.9 &  5.1 &  4.8&\multirow{2}{*}{$\&$}\\
  1600 &   5.3 &  \cellcolor[gray]{.8} 5.6 & \cellcolor[gray]{.8}  5.7 & \cellcolor[gray]{.8}  5.5 & \cellcolor[gray]{.8}  5.6 & \cellcolor[gray]{.8}  5.7 & \cellcolor[gray]{.8} 5.5 &  5.3 &  5.0 &  5.3 &  5.2 &  5.4& \multirow{3}{*}{$\frac{p(x)}{g_3(x)}$}\\
  3200 &  \cellcolor[gray]{.8} 6.1 & \cellcolor[gray]{.8}  5.9 &   5.1 & \cellcolor[gray]{.8}  5.5 &   5.4 & \cellcolor[gray]{.8}  5.6 & \cellcolor[gray]{.8} 5.5 &  5.4 &  5.4 &  4.8 &  4.9 &  4.9&\\
  6400 &  \cellcolor[gray]{.8} 5.9 &   5.4 & \cellcolor[gray]{.8}  5.6 & \cellcolor[gray]{.8}  5.8 &  \cellcolor[gray]{.8} 5.5 & \cellcolor[gray]{.8}  5.5 & \cellcolor[gray]{.8} 6.5 & \cellcolor[gray]{.8} 5.5 &  5.3 & \cellcolor[gray]{.8} 5.5 &  5.1 &  5.3&\\
\hline
   200 &   5.0 &   5.2 & \cellcolor[gray]{.8}  5.8 & \cellcolor[gray]{.8}  5.7 & \cellcolor[gray]{.8}  5.9 & \cellcolor[gray]{.8}  6.0 & \cellcolor[gray]{.8} 5.5 & \cellcolor[gray]{.8} 6.0 & \cellcolor[gray]{.8} 5.8 & \cellcolor[gray]{.8} 6.0 & \cellcolor[gray]{.8} 5.8 & \cellcolor[gray]{.8} 6.1&\multirow{3}{*}{$\frac{p(x)}{g_3(x)}$}\\
   400 &   5.3 & \cellcolor[gray]{.8}  5.5 & \cellcolor[gray]{.8}  5.6 & \cellcolor[gray]{.8}  5.5 & \cellcolor[gray]{.8}  5.8 &  \cellcolor[gray]{.8} 5.6 & \cellcolor[gray]{.8} 5.6 &  5.1 &  5.0 &  5.3 & \cellcolor[gray]{.8} 5.5 &  \cellcolor[gray]{.8} 5.5&\\
   800 &  \cellcolor[gray]{.8} 5.6 &   5.4 &   5.4 & \cellcolor[gray]{.8}  5.5 & \cellcolor[gray]{.8}  5.6 & \cellcolor[gray]{.8}  5.8 & \cellcolor[gray]{.8} 5.8 &  5.0 &  5.1 &  5.0 &  5.1 &  5.3&\multirow{2}{*}{$\&$}\\
  1600 &  \cellcolor[gray]{.8} 6.2 &   5.2 & \cellcolor[gray]{.8}  5.8 &   5.3 &   5.4 & \cellcolor[gray]{.8}  5.6 &  5.4 &  \cellcolor[gray]{.8}5.6 &  5.2 &  4.9 &  4.9 &  5.0&  \multirow{3}{*}{$\frac{p(x)}{g_3(x)}$}\\
  3200 &  \cellcolor[gray]{.8} 6.1 & \cellcolor[gray]{.8}  5.9 &   5.3 &  \cellcolor[gray]{.8} 5.9 &   4.9 &   5.2 & \cellcolor[gray]{.8} 5.9 & \cellcolor[gray]{.8} 5.6 &  5.2 &  5.1 &  4.8 &  5.4&\\
  6400 &  \cellcolor[gray]{.8} 6.1 & \cellcolor[gray]{.8}  5.5 & \cellcolor[gray]{.8}  6.0 &   \cellcolor[gray]{.8}5.5 &   4.7 &  \cellcolor[gray]{.8} 5.5 & \cellcolor[gray]{.8} 5.9 & \cellcolor[gray]{.8} 5.5 &  5.3 &  4.9 &  5.1 &  4.9&\\
\hline
\end{tabular}
\end{table}

\vspace *{0.0 cm}
\begin{table}[H]
\centering
\caption{ Sizes $\hat\alpha_s$  of  the test   $\hat {\check{X}}^2_{Med}$ (\ref{stdav2}) for comparison of two histograms with unnormalized weighted entries (left panel)  and sizes of the new test $\hat {\check{X}}^2$ (\ref{stddc5})  (right panel) for different pairs of weights (last column) and  numbers of events $n_1$, $n_2$.  Sizes of the tests that do not satisfy the hypothesis    $\alpha_s=5\%$     with a significance level equal to $0.05$   ($\hat\alpha_s>5.4\%$ or $\hat\alpha_s<4.6\%$) are highlighted with gray.}
\label{tab:result2}
\vspace *{0.2 cm}
\tiny
\begin{tabular}{r|rrrrrr||rrrrrr|c}
& \multicolumn{6}{c||}{$n_2$} & \multicolumn{6}{c|}{$n_2$} & \\
$n_1$&  200 &  400 &  800 & 1600 & 3200 & 6400 &  200 &  400 &  800 & 1600 & 3200 & 6400&$w(x)$\\
\hline
   200 & \cellcolor[gray]{.8}  5.8 & \cellcolor[gray]{.8}  5.6 &  \cellcolor[gray]{.8} 5.5 &  \cellcolor[gray]{.8} 6.2 &   \cellcolor[gray]{.8} 6.3 &  \cellcolor[gray]{.8} 5.8 &  4.9 &  5.1 &  5.0 &  5.0 &  5.2 & \cellcolor[gray]{.8} 5.6 &  \multirow{3}{*}{$\frac{2p(x)}{g_2(x)}$}\\
   400 & \cellcolor[gray]{.8}  5.6 &  \cellcolor[gray]{.8} 5.7 &  \cellcolor[gray]{.8} 6.0 &  \cellcolor[gray]{.8} 5.8 & \cellcolor[gray]{.8}  5.7 & \cellcolor[gray]{.8}  5.6 &  4.7 &  4.8 &  5.4 &  5.2 &  4.9 &  4.9 &\\
   800 & \cellcolor[gray]{.8}  6.0 & \cellcolor[gray]{.8}  6.1 & \cellcolor[gray]{.8}  5.8 &  \cellcolor[gray]{.8} 6.0 & \cellcolor[gray]{.8}  5.5 &  \cellcolor[gray]{.8} 6.1 &  5.2 &  5.2 &  5.0 &  5.1 &  5.1 &  4.8 &\multirow{2}{*}{$\&$}\\
  1600 &  \cellcolor[gray]{.8} 5.9 &  \cellcolor[gray]{.8} 5.6 &  \cellcolor[gray]{.8} 6.0 & \cellcolor[gray]{.8}  5.5 &   \cellcolor[gray]{.8}5.7 & \cellcolor[gray]{.8}  5.7 &  4.8 &  4.9 &  5.0 &  4.9 &  5.2 &  5.1&  \multirow{3}{*}{$\frac{3p(x)}{g_2(x)}$} \\
  3200 &  \cellcolor[gray]{.8} 5.9 & \cellcolor[gray]{.8}  5.5 &  \cellcolor[gray]{.8} 5.5 &  \cellcolor[gray]{.8} 5.9 &\cellcolor[gray]{.8}   5.6 &  \cellcolor[gray]{.8} 6.0 & \cellcolor[gray]{.8} 5.5 &  5.3 &  4.6 &  4.9 &  5.0 &  5.0&\\
  6400 &  \cellcolor[gray]{.8} 5.8 & \cellcolor[gray]{.8}  5.7 &  \cellcolor[gray]{.8} 5.8 &  \cellcolor[gray]{.8} 6.1 & \cellcolor[gray]{.8}  5.7 &   5.4 &  5.2 &  5.2 &  5.0 &  4.8 &  5.1 &  5.1&\\
\hline
   200 &   5.2 & \cellcolor[gray]{.8}  5.7 &  \cellcolor[gray]{.8} 5.9 & \cellcolor[gray]{.8}  5.7 & \cellcolor[gray]{.8}  5.7 & \cellcolor[gray]{.8}  5.9 &  5.0 &  5.1 &  4.6 &  5.0 &  4.9 &  5.4& \multirow{3}{*}{$\frac{2p(x)}{g_2(x)}$}\\
   400 &  \cellcolor[gray]{.8} 5.5 &   5.1 &   5.3 &  \cellcolor[gray]{.8} 5.9 & \cellcolor[gray]{.8}  5.6 & \cellcolor[gray]{.8}  5.9 &  5.1 & \cellcolor[gray]{.8} 5.7 &  5.2 &  5.3 &  5.1 &  4.9&\\
   800 &   5.2 &  \cellcolor[gray]{.8} 5.8 & \cellcolor[gray]{.8}  5.6 &  \cellcolor[gray]{.8} 5.8 &  \cellcolor[gray]{.8} 5.8 & \cellcolor[gray]{.8}  5.6 &  5.2 &  5.2 & 4.6 &  4.9 &  4.7 &  5.0&\multirow{2}{*}{$\&$}\\
  1600 &   5.3 &  \cellcolor[gray]{.8} 5.5 & \cellcolor[gray]{.8}  5.8 &  \cellcolor[gray]{.8} 5.7 &  \cellcolor[gray]{.8} 5.8 &  \cellcolor[gray]{.8} 5.9 &  5.3 &  4.9 &  5.0 &  5.4 & \cellcolor[gray]{.8} 4.4 &  4.7&   \multirow{3}{*}{$\frac{3p(x)}{g_3(x)}$} \\
  3200 & \cellcolor[gray]{.8}  5.5 &  \cellcolor[gray]{.8} 5.8 &   5.2 &   5.4 & \cellcolor[gray]{.8}  5.6 &  \cellcolor[gray]{.8} 6.1 &  4.6 &  5.3 &  5.1 &  5.1 &  4.6 &  5.2&\\
  6400 & \cellcolor[gray]{.8}  5.5 &   5.2 & \cellcolor[gray]{.8}  5.6 &  \cellcolor[gray]{.8} 6.1 &  \cellcolor[gray]{.8} 5.6 & \cellcolor[gray]{.8}  5.8 &  4.8 &  5.1 &  4.7 &  4.9 &  4.9 &  5.4&\\
\hline
   200 &   4.7 &   5.3 &   5.4 &   5.4 &  \cellcolor[gray]{.8} 5.6 &  \cellcolor[gray]{.8} 6.0 &  4.9 &  4.7 &  4.9 &  5.2 &  4.6 &  5.4&  \multirow{3}{*}{$\frac{2p(x)}{g_3(x)}$}\\
   400 &   5.0 &  \cellcolor[gray]{.8} 5.5 &  \cellcolor[gray]{.8} 5.8 &   5.4 & \cellcolor[gray]{.8}  5.6 &  5.5 &  5.0 &  4.8 &  5.3 &  5.2 &  5.4 &  5.2&\\
   800 &   5.4 &   5.2 &  \cellcolor[gray]{.8} 5.6 &  \cellcolor[gray]{.8} 5.5 &  \cellcolor[gray]{.8} 5.9 &  \cellcolor[gray]{.8} 5.7 &  4.9 &  5.2 &  5.0 &  5.1 &  5.1 &  5.4&\multirow{2}{*}{$\&$}\\
  1600 &  \cellcolor[gray]{.8} 5.6 &   5.2 &  \cellcolor[gray]{.8} 5.9 &   5.4 & \cellcolor[gray]{.8}  5.5 & \cellcolor[gray]{.8}  5.7 &  5.0 &  4.8 &  5.0 &  5.2 &  5.0 &  5.3&   \multirow{3}{*}{$\frac{3p(x)}{g_3(x)}$} \\
  3200 &  \cellcolor[gray]{.8} 6.0 &  \cellcolor[gray]{.8} 5.9 &   5.4 & \cellcolor[gray]{.8}  6.0 &   5.3 &   5.4 &  5.2 &  5.3 &  5.0 &  5.0 &  5.1 &  5.0&\\
  6400 &  \cellcolor[gray]{.8} 5.8 &  \cellcolor[gray]{.8} 5.5 &  \cellcolor[gray]{.8} 6.0 & \cellcolor[gray]{.8}  5.9 &   5.1 & \cellcolor[gray]{.8}  5.6 & \cellcolor[gray]{.8} 5.7 &  5.3 &  5.1 &  4.7 &  5.1 &  4.7&\\
\hline
\end{tabular}
\end{table}

\vspace *{-2.4 cm}
\begin{table}[H]
\centering
\caption{ Sizes $\hat\alpha_s$  of the test   $_1\hat {\check{X}}^2_{Med}$ (\ref{stdav2}) for comparison of two histograms  with normalized  and unnormalized weighted entries  (left panel) and sizes of  the  new test $_1\hat {\check{X}}^2$ (\ref{stddc5})  (right panel) for different pairs of weights (last column) and numbers of events $n_1$, $n_2$. Sizes of the tests that do not satisfy the hypothesis  $\alpha_s=5\%$ ($\hat\alpha_s>5.4\%$ or $\hat\alpha_s<4.6\%$) with a significance level equal to $0.05$ are highlighted with gray.}
\label{tab:result3}
\vspace *{0.2 cm}
\scriptsize
\tiny
\begin{tabular}{r|rrrrrr||rrrrrr|c}
& \multicolumn{6}{c||}{$n_2$} & \multicolumn{6}{c|}{$n_2$} & \\
$n_1$&  200 &  400 &  800 & 1600 & 3200 & 6400 &  200 &  400 &  800 & 1600 & 3200 & 6400&$w(x)$\\
\hline
   200 &   5.4 &  \cellcolor[gray]{.8} 5.6 &   5.2 &   5.2 &   5.3 & \cellcolor[gray]{.8}  5.5 &  4.9 &  4.8 &  4.9 &  5.1 &  4.8 &  4.7&\multirow{3}{*}{$1$}\\
   400 & \cellcolor[gray]{.8}  5.7 &  \cellcolor[gray]{.8} 5.8 &   5.4 & \cellcolor[gray]{.8}  5.9 &   5.3 &  \cellcolor[gray]{.8} 5.6&  5.3 &  5.0 &  5.0 &  4.8 &  5.1 &  5.0&\\
   800 & \cellcolor[gray]{.8}  5.6 &  \cellcolor[gray]{.8} 6.0 &  \cellcolor[gray]{.8} 5.8 &  \cellcolor[gray]{.8} 5.9 &   5.2 &  \cellcolor[gray]{.8} 5.5 &  5.1 &  4.9 & 4.6 &  5.2 &  5.1 &  5.1   &\multirow{2}{*}{$\&$}\\
  1600 &   5.4 &  \cellcolor[gray]{.8} 5.7 & \cellcolor[gray]{.8}  6.0 &   5.4 &  \cellcolor[gray]{.8} 5.7 &   5.4 &  5.3 &  5.3 &  4.7 &  5.1 &  4.9 &  5.0&   \multirow{3}{*}{$\frac{3p(x)}{g_2(x)}$}\\
  3200 &  \cellcolor[gray]{.8} 5.7 &  \cellcolor[gray]{.8} 5.7 & \cellcolor[gray]{.8}  5.7 &  \cellcolor[gray]{.8} 6.0 & \cellcolor[gray]{.8}  5.5 &  \cellcolor[gray]{.8} 5.5 &  5.3 &  5.0 &  4.9 &  5.2 &  4.8 &  5.0&\\
  6400 &  \cellcolor[gray]{.8} 5.7 &  \cellcolor[gray]{.8} 5.8 & \cellcolor[gray]{.8}  5.7 & \cellcolor[gray]{.8}  5.5 & \cellcolor[gray]{.8}  5.6 & \cellcolor[gray]{.8}  5.6 &  5.0 &  5.1 &  4.6 & \cellcolor[gray]{.8} 4.5 &  5.2 &  5.2&\\
\hline
   200 &   4.8 &   4.8 &   5.1 &   4.9 &   5.2 & \cellcolor[gray]{.8}  5.6 &  4.9 &  5.3 &  5.2 &  5.0 &  5.1 &  4.9& \multirow{3}{*}{$1$}\\
   400 &   5.2 &   5.2 &   5.1 &   5.3 & \cellcolor[gray]{.8}  5.5 &  \cellcolor[gray]{.8} 5.9 &  5.1 &  4.9 &  4.8 &  5.0 &  5.2 &  4.8&\\
   800 &   5.4 &  \cellcolor[gray]{.8} 5.8 &   5.3 &  \cellcolor[gray]{.8} 5.8 &   5.3 &   5.3 &  4.6 &  4.9 &  4.8 &  5.0 &  4.8 &  4.8& \multirow{2}{*}{$\&$}\\
  1600 &   5.3 & \cellcolor[gray]{.8}  5.6 &  \cellcolor[gray]{.8} 5.8 &   5.4 &  \cellcolor[gray]{.8} 6.0 &  \cellcolor[gray]{.8} 5.6 &  4.7 &  5.0 &  5.1 &  4.9 &  4.8 &  5.0&  \multirow{3}{*}{$\frac{3p(x)}{g_3(x)}$} \\
  3200 &  \cellcolor[gray]{.8} 5.6 &   5.2 &   5.4 &  \cellcolor[gray]{.8} 5.7 & \cellcolor[gray]{.8}  5.5 &   5.4 &  5.0 &  4.9 &  4.9 &  4.9 &  4.8 &  4.7&\\
  6400 & \cellcolor[gray]{.8}  5.8 &  \cellcolor[gray]{.8} 5.5 &  \cellcolor[gray]{.8} 5.6 &  \cellcolor[gray]{.8} 5.5 & \cellcolor[gray]{.8}  5.9 & \cellcolor[gray]{.8}  5.6 &  5.1 &  5.1 & 4.6 &  4.9 &  5.2 &  4.8&\\
\hline
   200 &  \cellcolor[gray]{.8} 6.1 &  \cellcolor[gray]{.8} 5.8 & \cellcolor[gray]{.8}  5.6 &  \cellcolor[gray]{.8} 6.3 &  \cellcolor[gray]{.8} 6.5 & \cellcolor[gray]{.8}  6.1 & \cellcolor[gray]{.8} 6.1 & \cellcolor[gray]{.8} 5.6 & \cellcolor[gray]{.8} 5.9 & \cellcolor[gray]{.8} 5.6 & \cellcolor[gray]{.8} 6.1 & \cellcolor[gray]{.8} 5.9&\multirow{3}{*}{$\frac{p(x)}{g_2(x)}$}\\
   400 & \cellcolor[gray]{.8}  5.7 &  \cellcolor[gray]{.8} 5.9 & \cellcolor[gray]{.8}  6.2 &  \cellcolor[gray]{.8} 5.9 & \cellcolor[gray]{.8}  5.8 & \cellcolor[gray]{.8}  5.6 & \cellcolor[gray]{.8} 5.8 & \cellcolor[gray]{.8} 5.8 &  5.1 &  5.3 &  5.4 &  5.1&\\
   800 &  \cellcolor[gray]{.8} 6.0 &  \cellcolor[gray]{.8} 6.2 &  \cellcolor[gray]{.8} 5.8 &  \cellcolor[gray]{.8} 6.0 &\cellcolor[gray]{.8}   5.6 &  \cellcolor[gray]{.8} 6.1 & \cellcolor[gray]{.8} 6.1 &  5.3 &  4.9 &  5.2 &  5.0 &  5.1&\multirow{2}{*}{$\&$}\\
  1600 &  \cellcolor[gray]{.8} 5.9 & \cellcolor[gray]{.8}  5.6 & \cellcolor[gray]{.8}  6.1 & \cellcolor[gray]{.8}  5.5 & \cellcolor[gray]{.8}  5.7 & \cellcolor[gray]{.8}  5.7 &  5.2 &  5.1 &  4.8 &  5.0 &  5.2 &  5.0&  \multirow{3}{*}{$\frac{3p(x)}{g_2(x)}$} \\
  3200 & \cellcolor[gray]{.8}  5.9 & \cellcolor[gray]{.8}  5.5 & \cellcolor[gray]{.8}  5.5 & \cellcolor[gray]{.8}  6.0 & \cellcolor[gray]{.8}  5.6 &  \cellcolor[gray]{.8} 6.0 &  4.8 &  5.0 &  5.2 &  5.2 &  5.1 &  5.2&\\
  6400 &  \cellcolor[gray]{.8} 5.8 &  \cellcolor[gray]{.8} 5.7 &  \cellcolor[gray]{.8} 5.8 &  \cellcolor[gray]{.8} 6.1 & \cellcolor[gray]{.8}  5.7 &   5.4 &  5.1 &  5.4 &  5.2 &  5.2 &  5.2 & \cellcolor[gray]{.8} 5.7&\\
\hline
   200 &   5.4 &  \cellcolor[gray]{.8} 5.9 &  \cellcolor[gray]{.8} 6.0 & \cellcolor[gray]{.8}  5.8 &  \cellcolor[gray]{.8} 5.8 &  \cellcolor[gray]{.8} 6.1 & \cellcolor[gray]{.8} 5.9 & \cellcolor[gray]{.8} 6.3 & \cellcolor[gray]{.8} 5.8 & \cellcolor[gray]{.8} 6.3 & \cellcolor[gray]{.8} 5.8 & \cellcolor[gray]{.8} 5.7&\multirow{3}{*}{$\frac{p(x)}{g_2(x)}$}\\
   400 & \cellcolor[gray]{.8}  5.6 &   5.2 &   5.4 &  \cellcolor[gray]{.8} 6.1 & \cellcolor[gray]{.8}  5.7 & \cellcolor[gray]{.8}  5.9 & \cellcolor[gray]{.8} 5.8 & \cellcolor[gray]{.8} 5.6 &  5.4 &  5.4 &  5.3 &  5.4&\\
   800 &   5.3 & \cellcolor[gray]{.8}  5.9 & \cellcolor[gray]{.8}  5.7 &  \cellcolor[gray]{.8} 5.8 & \cellcolor[gray]{.8}  5.9 & \cellcolor[gray]{.8}  5.6 &  5.4 &  5.1 &  5.4 &  5.4 &  5.2 &  5.2&\multirow{2}{*}{$\&$}\\
  1600 &   5.4 & \cellcolor[gray]{.8}  5.5 &  \cellcolor[gray]{.8} 5.8 & \cellcolor[gray]{.8}  5.8 & \cellcolor[gray]{.8}  5.8 &   \cellcolor[gray]{.8} 5.9 & \cellcolor[gray]{.8} 5.9 &  5.0 &  5.0 &  4.8 &  4.6 &  4.8& \multirow{3}{*}{$\frac{3p(x)}{g_3(x)}$} \\
  3200 & \cellcolor[gray]{.8}  5.5 &   \cellcolor[gray]{.8}5.8 &   5.2 &   5.4 & \cellcolor[gray]{.8}  5.6 & \cellcolor[gray]{.8}  6.1 &  5.0 &  4.8 &  4.9 &  5.1 &  4.8 &  5.1&\\
  6400 &  \cellcolor[gray]{.8} 5.5 &   5.2 &  \cellcolor[gray]{.8} 5.6 & \cellcolor[gray]{.8}  6.1 &  \cellcolor[gray]{.8} 5.6 & \cellcolor[gray]{.8}  5.8 &  5.2 &  4.9 &  5.0 &  4.6 &  5.4 &  5.4&\\
\hline
  200 &  \cellcolor[gray]{.8} 5.7 & \cellcolor[gray]{.8}\cellcolor[gray]{.8}  5.6 &  \cellcolor[gray]{.8} 5.8 & \cellcolor[gray]{.8}  5.8 &  \cellcolor[gray]{.8} 5.7 &  \cellcolor[gray]{.8} 5.6 & \cellcolor[gray]{.8} 5.8 & \cellcolor[gray]{.8} 6.0 & \cellcolor[gray]{.8} 5.7 & \cellcolor[gray]{.8} 5.7 & \cellcolor[gray]{.8} 5.5 & \cellcolor[gray]{.8} 5.5& \multirow{3}{*}{$\frac{p(x)}{g_3(x)}$}\\
   400 &  \cellcolor[gray]{.8} 6.1 &   \cellcolor[gray]{.8} 5.7 &   5.2 &  \cellcolor[gray]{.8} 6.2 &  \cellcolor[gray]{.8} 5.6 &  \cellcolor[gray]{.8} 5.5&  5.2 &  5.3 & \cellcolor[gray]{.8} 5.5 & \cellcolor[gray]{.8} 5.6 &  5.4 &  5.1&\\
   800 &  \cellcolor[gray]{.8} 5.7 &  \cellcolor[gray]{.8} 5.8 & \cellcolor[gray]{.8}  5.9 &  \cellcolor[gray]{.8} 6.0 & \cellcolor[gray]{.8}  5.5 &\cellcolor[gray]{.8}   5.5 &  5.3 &  5.4 &  5.3 &  5.0 &  5.0 &  5.1&\multirow{2}{*}{$\&$}\\
  1600 & \cellcolor[gray]{.8}  5.8 &  \cellcolor[gray]{.8} 5.5 & \cellcolor[gray]{.8}  5.9 &  \cellcolor[gray]{.8} 5.6 & \cellcolor[gray]{.8}  5.7 &  \cellcolor[gray]{.8} 5.8 &  4.8 &  5.1 &  4.9 &  5.0 &  5.2 &  4.9&  \multirow{3}{*}{$\frac{3p(x)}{g_2(x)}$}\\
  3200 &  \cellcolor[gray]{.8} 5.6 &  \cellcolor[gray]{.8} 5.9 & \cellcolor[gray]{.8}  5.6 &  \cellcolor[gray]{.8} 5.8 & \cellcolor[gray]{.8}  5.5 & \cellcolor[gray]{.8}  5.6 &  5.3 &  5.3 &  5.1 &  4.9 &  5.1 &  4.8&\\
  6400 & \cellcolor[gray]{.8}  5.9 &  \cellcolor[gray]{.8} 5.5 &  \cellcolor[gray]{.8} 5.8 &  \cellcolor[gray]{.8} 5.9 & \cellcolor[gray]{.8}  5.6 &  \cellcolor[gray]{.8} 6.1 &  5.1 &  4.9 &  5.2 &  4.8 &  4.9 &  5.1&\\
\hline
   200 &   5.0 &  \cellcolor[gray]{.8} 5.5 &  \cellcolor[gray]{.8} 5.6 & \cellcolor[gray]{.8}  5.5 & \cellcolor[gray]{.8}  5.6 &  \cellcolor[gray]{.8} 6.0& \cellcolor[gray]{.8} 5.7 & \cellcolor[gray]{.8} 6.1 &  \cellcolor[gray]{.8} 5.9 & \cellcolor[gray]{.8} 5.8 & \cellcolor[gray]{.8} 5.6 &  \cellcolor[gray]{.8}5.5&\multirow{3}{*}{$\frac{p(x)}{g_3(x)}$}\\
   400 &   5.2 & \cellcolor[gray]{.8}  5.6 &  \cellcolor[gray]{.8} 5.9 &  \cellcolor[gray]{.8} 5.6 &  \cellcolor[gray]{.8} 5.7 &  \cellcolor[gray]{.8} 5.6 &\cellcolor[gray]{.8}  5.7 &  5.2 &  5.2 &  5.3 & \cellcolor[gray]{.8} 5.7 &  5.2&\\
   800 &  \cellcolor[gray]{.8} 5.5 &   5.4 &  \cellcolor[gray]{.8} 5.6 &   \cellcolor[gray]{.8}5.5 &  \cellcolor[gray]{.8} 5.9 &  \cellcolor[gray]{.8} 5.8 &  5.3 &  5.0 &  5.2 &  4.8 &  5.0 &  5.4&\multirow{2}{*}{$\&$}\\
  1600 & \cellcolor[gray]{.8}  5.6 &   5.2 &  \cellcolor[gray]{.8} 6.0 &   5.4 &  \cellcolor[gray]{.8} 5.6 & \cellcolor[gray]{.8}  5.7 &  5.1 &  5.0 &  5.0 &  5.3 &  5.0 &  4.8& \multirow{3}{*}{$\frac{3p(x)}{g_3(x)}$}  \\
  3200 & \cellcolor[gray]{.8}  6.0 &  \cellcolor[gray]{.8} 5.9 &  \cellcolor[gray]{.8} 5.5 &  \cellcolor[gray]{.8} 6.0 &   5.3 &   5.4 &  5.3 &  5.0 &  4.8 &  4.8 &  5.3 &  5.1&\\
  6400 &  \cellcolor[gray]{.8} 5.8 &  \cellcolor[gray]{.8} 5.5 &  \cellcolor[gray]{.8} 6.0 &  \cellcolor[gray]{.8} 5.9 &   5.1 &  \cellcolor[gray]{.8} 5.6 &  5.1 &  5.3 &  4.9 &  5.0 &  5.1 &  4.9&\\
\hline
\end{tabular}
\end{table}

 Distributions  of p-value were studied  by simulating 100\,000 runs.  In each run 3\,200 events  were simulated for one histogram and 6\,400 events for another one.  Distributions were calculated for:   
\begin{itemize}
\item  The  median statistic $\hat X^2_{Med}$ (\ref{stdavu})  and the new statistic $\hat X^2$ (\ref{stdd3}) used for comparison of two  histograms with normalized weights.   The first histogram  represents the PDF $p(x)$  with weights of events  $\frac{p(x)}{g_2(x)}$  and  the second  histogram  represents the  PDF $p(x)$  with weights of events $\frac{p(x)}{g_3(x)}$;
\item  The median statistic  $\hat {\check{X}}^2_{Med}$ (\ref{stdav2})  and the new statistic $\hat {\check{X}}^2$ (\ref{stddc5}) used for comparison of  two  histograms with unnormalized weights. The  first  histogram  represents the  PDF  $p(x)$  with weights  $\frac{2p(x)}{g_2(x)}$  and  the second histogram represents the PDF  $p(x)$ with weights of events $\frac{3p(x)}{g_3(x)}$.
\end {itemize}    
\begin{figure}[H]
\includegraphics[width=1.05 \textwidth]{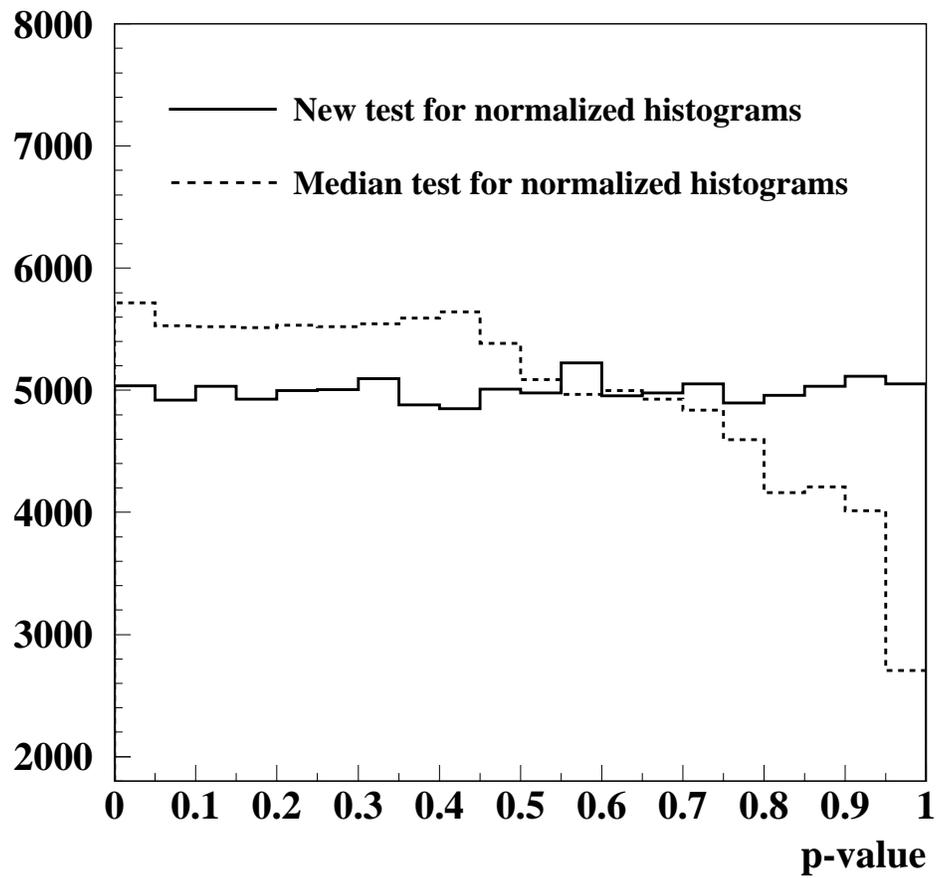}\\
\vspace*{-1. cm}
 \caption {Distributions of p-value for the median statistics $\hat X^2_{Med}$ (\ref{stdavu})   and the new statistic $\hat X^2$ (\ref{stdd3}) used for comparison of  two weighted histograms with normalized weights.}
\vspace*{1 cm }
\end{figure}

\begin{figure}[H]
\includegraphics[width=1.05\textwidth]{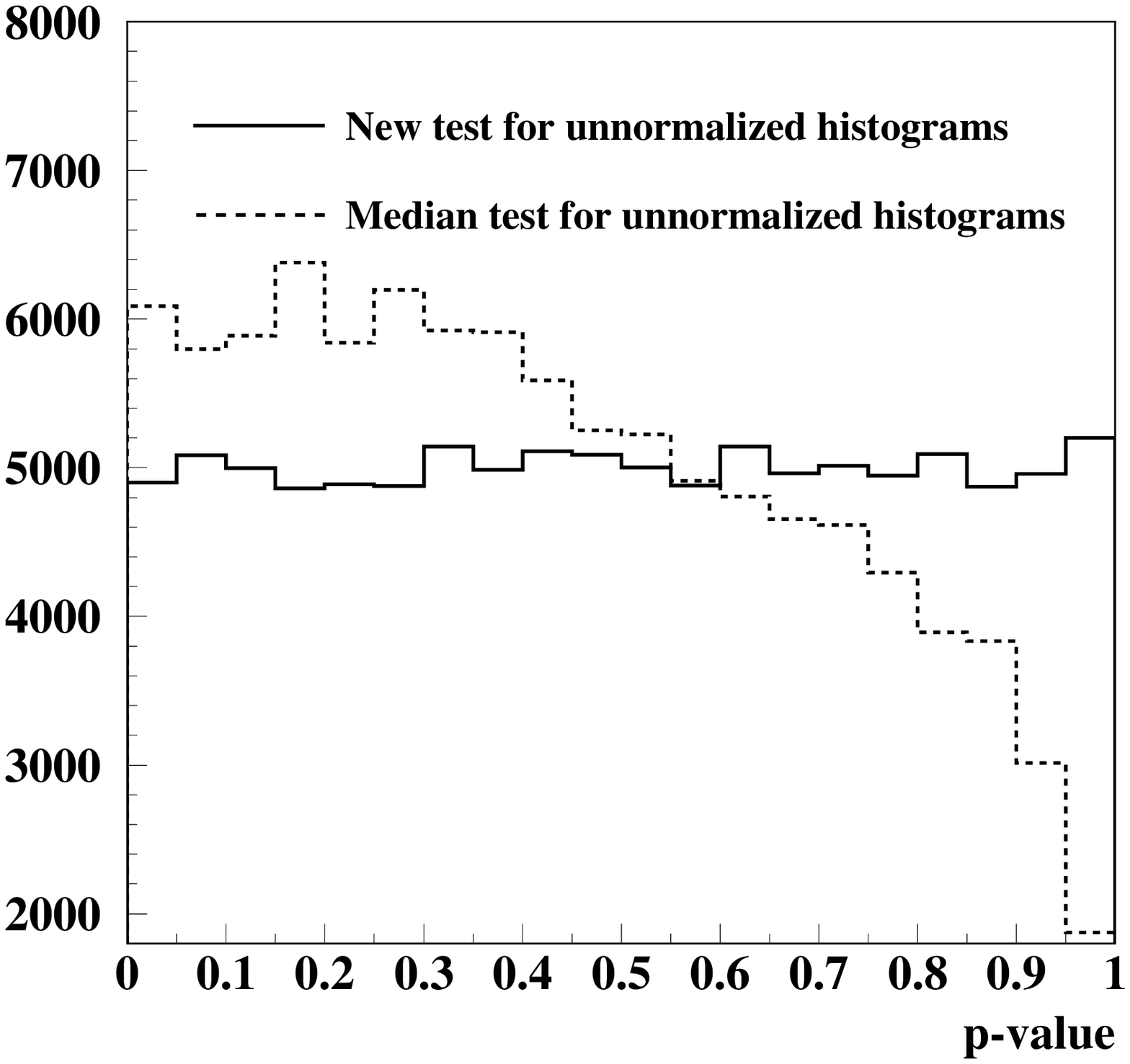}\\
\vspace*{-1 cm}
 \caption {Distributions of p-value for the median statistics   $\hat {\check{X}}^2_{Med}$ (\ref{stdav2}) and the new statistic   $\hat {\check{X}}^2$ (\ref{stddc5}) used for comparison of  two weighted histograms with unnormalized weights.}
\vspace*{1 cm }
\end{figure}
\newpage
\noindent
{\it Conclusions to subsection 6.1}
\begin{itemize}
\item Tables 1-3\\
 The sizes $\hat\alpha_s$  of the new tests $\hat X^2$ (\ref{stdd3}), $\hat {\check{X}}^2$, $_1\hat {\check{X}}^2$ (\ref{stddc5})  are closer to a nominal value of a test size equal to 5\%,  than the sizes of the  median statistics $\hat X^2_{Med}$(\ref{stdavu}), $\hat {\check{X}}^2_{Med}$, $_1\hat {\check{X}}^2_{Med}$ (\ref{stdav2}).
\item Figure 3\\
The distribution of the new statistic $\hat X^2$ (\ref{stdd3}) for comparison of weighted histograms with normalized weights  is  closer to a $\chi^2_{m-1}$ distribution, than the distribution  of the median statistic  $\hat X^2_{Med}$ (\ref{stdavu}). 
\item Figure 4\\
The distribution of the new statistic $\hat {\check{X}}^2$ (\ref{stddc5})  for comparison of weighted histograms with unnormalized weights is  closer to  a  $\chi^2_{m-2}$ distribution, than the distribution of the median statistics $\hat {\check{X}}^2_{Med}$ (\ref{stdav2}).
\end{itemize}

\subsection{Tests for histograms with Poisson distribution of events}

The size and power  of the tests were calculated for a different 
number of events  defined by  parameters $n_{01}$ and  $n_{02}$ of a Poisson distribution   in five bin histograms. 
 In the following, numerical examples demonstrate application of: 
\begin{itemize}
  \item The   test $X^2_{p2}$ (\ref{pois2}) and  the new  test  $\hat X^2$ (\ref{stdd3}) for comparison of weighted histograms with normalized
weights  (Tables~\ref{tab:result1c} and  \ref{tab:result1pw});
  \item The  test $X^2_{p2}$ (\ref{pois2}) and the new test $\hat {\check{X}}^2$ (\ref{stddc5}) for comparison of weighted histograms with unnormalized weights (Tables~\ref{tab:result2c} and \ref{tab:result2pw});
  \item  The  test $X^2_{p2}$ (\ref{pois2})   and the  new test  $_1\hat {\check{X}}^2 $(\ref{stddc5}) for comparison of weighted histogram with normalized weights and the histogram with unnormalized weights (Tables~\ref{tab:result3c} and \ref{tab:result3pw}). 

\end{itemize}
\vspace *{-4cm}
\subsubsection {Size of tests for  the Poisson weighted histograms }
A total number of events  for  histograms  was simulated according to a Poisson distribution  with parameters $n_{01}$ and  $n_{02}$.
\vspace *{-0.5cm}
\begin{table}[H]
\centering
\caption{ Sizes $\hat\alpha_s$  of  the test $X^2_{p2}$ (\ref{pois2}) for comparison of two Poisson histograms with normalized weights (left panel) and sizes of the new test  $\hat X^2$ (\ref{stdd3})  (right panel) for different pairs of weights (last column) and  parameters  $n_{01}$, $n_{02}$. Sizes of the tests that do not satisfy the hypothesis $\alpha_s=5\%$ with a significance level equal to $0.05$ ($\hat\alpha_s>5.4\%$ or $\hat\alpha_s<4.6\%$) are highlighted with gray.}
\label{tab:result1c}
\vspace *{0.2 cm}
\tiny
\begin{tabular}{r|rrrrrr||rrrrrr|c}
& \multicolumn{6}{c||}{$n_{02}$} & \multicolumn{6}{c|}{$n_{02}$} & \\
$n_{01}$&  200 &  400 &  800 & 1600 & 3200 & 6400 &  200 &  400 &  800 & 1600 & 3200 & 6400&$w(x)$\\
\hline
 200 &  4.8 &  5.2 &  5.2 &  4.8 &  4.9 &  4.8& \cellcolor[gray]{.8} 4.5 & \cellcolor[gray]{.8} 4.4 &  4.8 &  4.8 &  5.2 &  5.3& \multirow{3}{*}{$1$}\\
   400 &  4.7 &  4.7 &  4.8 &  5.1 &  4.8 &  4.8& \cellcolor[gray]{.8} 4.2 & \cellcolor[gray]{.8} 4.3 &  4.8 &  4.9 &  4.7 &  4.9&\\
   800 &  4.6 &  4.8 &   \cellcolor[gray]{.8}4.5 &  5.2 &  5.2 &  4.9 &  4.7 & \cellcolor[gray]{.8} 4.5 & \cellcolor[gray]{.8} 4.4 &  5.1 &  5.0 &  5.0& \multirow{2}{*}{$\&$}\\
  1600 &  5.1 &  4.9 &  5.1 & \cellcolor[gray]{.8} 4.4 &  4.6 &  4.9 &  4.9 &  4.6 &  4.7 &  4.9 &  4.9 &  4.9& \multirow{3}{*}{$1$}\\
  3200 & \cellcolor[gray]{.8} 4.5 &  4.8 &  5.0 &  4.6 &  5.0 &  5.0 &  4.8 &  4.8 &  4.9 &  4.8 &  4.8 &  4.9&\\
  6400 &  4.9 & \cellcolor[gray]{.8} 4.5 &  5.3 &  5.2 &  4.9 &  4.7 &  5.1 &  4.9 &  5.2 &  5.0 &  4.8 &  4.8&\\
   \hline
   200 &  4.7 &  5.2 &\cellcolor[gray]{.8}  4.4 &  4.7 &  4.8 &  \cellcolor[gray]{.8}4.3 &  5.1 &  5.0 &  4.8 &  4.8 &  4.8 & \cellcolor[gray]{.8} 4.4& \multirow{3}{*}{$1$}\\
   400 &  5.1 &  5.2 &  4.9 &  4.7 &  4.9 &  5.2 &  5.2 &  5.1 &  4.7 &  5.3 &  5.3 &  4.8&\\
   800 &  5.2 &  5.2 &  5.0 &  5.0 &  4.6 &  4.9 &  5.2 &  5.1 &  4.9 &  5.1 &  5.0 & \cellcolor[gray]{.8} 4.3&\multirow{2}{*}{$\&$}\\
  1600 &  4.9 &  5.3 &  4.7 &  5.1 &  5.1 &  4.7 &  5.3 &  5.4 &  5.0 &  4.9 &  4.7 &  5.2& \multirow{3}{*}{$\frac{p(x)}{g_2(x)}$}\\
  3200 &  5.2 &  4.9 &  5.0 &  4.7 &  4.9 &  5.2 &  \cellcolor[gray]{.8}5.5 &  5.1 &  5.1 &  5.1 &  5.2 &  5.2&\\
  6400 &  4.9 &  5.2 & \cellcolor[gray]{.8} 4.4 &  4.7 &  5.0 &  4.6 &  5.4 &  5.3 &  4.9 &  5.2 &  5.2 &  4.9&\\
   \hline
   200 &  4.7 &  4.7 &  4.8 &  4.9 &  4.7 &  4.6 &  5.2 &  4.6 &  5.1 &  4.6 &  5.0 &  4.9&\multirow{2}{*}{$1$}
   \\
   400 &  5.1 &  5.1 &  4.6 &  4.9 &  4.5 &  4.6 &  5.1 &  4.6 &  5.2 &  5.2 &  4.8 &  4.9& \\
   800 &  4.9 &  4.9 &  5.3 &  5.0 &  4.8 &  5.3 & \cellcolor[gray]{.8} 5.5 &  5.1 &  5.2 &  5.0 &  4.7 &  5.3&\multirow{2}{*}{$\&$}\\
  1600 &  5.2 & \cellcolor[gray]{.8} 5.7 &  5.0 &  5.2 &  5.0 &  5.1 & \cellcolor[gray]{.8} 5.8 &  5.2 &  4.9 &  5.0 &  5.0 &  5.0& \multirow{3}{*}{$\frac{p(x)}{g_3(x)}$} \\
  3200 &  5.2 &  5.0 &  4.8 &  5.1 &  5.0 &  4.9 & \cellcolor[gray]{.8} 6.0 &  5.3 &  5.0 &  5.2 &  5.1 &  4.9&\\
  6400 &  5.1 &  4.9 &  4.7 &  4.9 &  5.0 &  5.4 & \cellcolor[gray]{.8} 5.9 &  5.4 & \cellcolor[gray]{.8} 5.6 &  4.9 &  5.1 &  4.7&\\
   \hline
   200 &  5.3 &  4.9 &  4.9 &  5.0 &  5.0 &  4.8&  \cellcolor[gray]{.8}5.5 &  5.3 &  5.3 &  \cellcolor[gray]{.8}5.5 &  5.2 & \cellcolor[gray]{.8} 5.6& \multirow{3}{*}{$\frac{p(x)}{g_2(x)}$}\\
   400 &  5.1 &  4.6 &  5.0 &  4.8 &  4.9 &  4.7 &  5.1 &  5.2 &  5.4 &  4.8 &  5.4 &  5.3&\\
   800 &  5.2 &  4.8 &  5.1 &  5.2 &  4.9 &  5.0 &  5.3 &  5.3 &  5.2 &  4.8 &  5.0 &  4.7&\multirow{2}{*}{$\&$}\\
  1600 &  5.3 &  5.1 &  5.1 &  5.1 &  5.0 &  4.8 &  5.3 &  5.2 &  5.1 & \cellcolor[gray]{.8} 5.5 &  5.3 &  4.9&  \multirow{3}{*}{$\frac{p(x)}{g_2(x)}$}\\
  3200 &  5.1 &  4.8 &  5.4 &  4.8 &  4.9 &  5.1 &  5.2 &  5.4 &  5.2 &  5.1 &  5.4 &  5.1&\\
  6400 &  5.3 &  5.2 &  5.1 &  5.2 &  5.0 &  5.1 &  \cellcolor[gray]{.8}5.5 & \cellcolor[gray]{.8} 5.7 &  5.2 &  5.2 &  5.3 &  5.3&\\
   \hline
   200 &  4.8 &  5.0 &  5.2 &  5.3 &  4.8 &  5.1 &  5.1 &  5.3 &  5.3 &  5.3 &  5.1 &  \cellcolor[gray]{.8}5.5& \multirow{3}{*}{$\frac{p(x)}{g_2(x)}$}\\
   400 &  4.9 &  5.3 &  5.1 &  5.0 &  4.8 &  5.1 & \cellcolor[gray]{.8} 5.5 &  5.3 &  4.9 & \cellcolor[gray]{.8} 5.6 &  4.9 &  4.9&\\
   800 &  4.8 &  5.0 &  5.0 &  4.9 &  5.3 &  4.6 &  4.9 &  5.1 &  5.0 &  5.4 &  5.2 &  4.9&\multirow{2}{*}{$\&$}\\
  1600 &  4.8 &  4.8 &  5.1 &  4.9 &  5.2 &  5.1&\cellcolor[gray]{.8}  5.6 &  5.2 &  5.1 & \cellcolor[gray]{.8} 5.5 & \cellcolor[gray]{.8} 5.7 &  5.2& \multirow{3}{*}{$\frac{p(x)}{g_3(x)}$}\\
  3200 &  4.9 &  5.3 &  5.1 &  4.9 &  5.0 &  4.8 & \cellcolor[gray]{.8} 5.7 &  5.1 &  5.1 &  4.9 &  4.6 &  5.3&\\
  6400 &  4.9 &  5.1 &  4.9 &  5.1 &  5.3 &  5.0 &\cellcolor[gray]{.8}  5.7 &  5.3 &  4.9 &  5.4 &  5.1 &  5.2&\\
  \hline
   200 &  5.1 &  4.8 &  4.9 &  5.0 &  4.8 & \cellcolor[gray]{.8} 5.5 &  5.4 & \cellcolor[gray]{.8} 5.9 & \cellcolor[gray]{.8} 5.6 & \cellcolor[gray]{.8} 5.6 & \cellcolor[gray]{.8} 6.2 & \cellcolor[gray]{.8} 5.9& \multirow{3}{*}{$\frac{p(x)}{g_3(x)}$}\\
   400 &  5.1 &  4.9 &  5.0 &  5.3 &  4.9 &  5.0 & \cellcolor[gray]{.8} 5.7 &  5.3 &  5.4 &  5.4 & \cellcolor[gray]{.8} 5.5 & \cellcolor[gray]{.8} 5.5&\\
   800 &  5.0 &  5.0 &  5.4 &  4.7 &  4.7 &  4.9 & \cellcolor[gray]{.8} 5.6 &  4.9 &  5.0 &  5.1 &  5.2 &  5.2&\multirow{2}{*}{$\&$}\\
  1600 &  5.1 &  5.0 &  4.9 &  4.7 &  4.8 &  4.9 & \cellcolor[gray]{.8} 5.5 &  5.4 &  5.3 &  5.1 &  4.9 &  4.7& \multirow{3}{*}{$\frac{p(x)}{g_3(x)}$}\\
  3200 &  4.7 &  4.9 &  5.1 &  4.9 &  4.9 &  5.2 & \cellcolor[gray]{.8} 6.0 &  5.0 &  5.0 &  4.9 &  4.9 &  5.3&\\
  6400 &  5.1 &  \cellcolor[gray]{.8} 4.5 &  5.0 &  4.9 &  4.8 &  5.2 & \cellcolor[gray]{.8} 5.8 &  5.3 &  5.1 &  4.9 &  5.3 &  4.8&\\
 \hline
 \end{tabular}
\end{table}
\vspace *{-0.0cm}
\begin{table}[H]
\centering
\caption{Sizes $\hat\alpha_s$  of the  test $X^2_{p2}$ (\ref{pois}) for comparison of two Poisson histograms with unnormalized weighted entries (left panel)  and sizes of the  new test  $\hat {\check{X}}^2$(\ref{stddc5})  (right panel) for different pairs of weights (last column)  and parameters  $n_{01}$, $n_{02}$.   Sizes of the tests that do not satisfy the hypothesis    $\alpha_s=5\%$     with a significance level equal to $0.05$   ($\hat\alpha_s>5.4\%$ or $\hat\alpha_s<4.6\%$) are highlighted with gray.}
\label{tab:result2c}
\vspace *{0.2 cm}
\tiny
\begin{tabular}{r|rrrrrr||rrrrrr|c}
& \multicolumn{6}{c||}{$n_{02}$} & \multicolumn{6}{c|}{$n_{02}$} & \\
$n_{01}$&  200 &  400 &  800 & 1600 & 3200 & 6400 &  200 &  400 &  800 & 1600 & 3200 & 6400&$w(x)$\\
\hline
 200 &  5.1 &  5.2 &  5.3 &  5.3 &  4.9 &  5.0 &  4.8 &  4.8 &  4.9 &  5.2 &  5.0 &  5.2& \multirow{3}{*}{$\frac{2p(x)}{g_2(x)}$}\\
   400 &  4.7 &  5.4 &  4.8 &  5.0 &  5.1 &  4.8 &  4.9 &  5.2 &  4.8 &  5.0 &  5.3 &  4.9&\\
   800 &  4.8 &  4.8 &  4.8 &  5.3 &  4.8 &  5.0 &  4.6 &  5.0 &  4.6 &  5.3 &  4.7 &  5.2& \multirow{2}{*}{$\&$}\\
  1600 &  5.0 &  5.0 &  5.3 &  4.8 &  5.1 &  4.7 &  4.8 &  4.8 &  5.2 &  4.8 &  4.9 &  4.9& \multirow{3}{*}{$\frac{3p(x)}{g_2(x)}$}\\
  3200 &  5.1 &  5.2 &  5.1 &  5.0 &  4.9 &  5.0 &  4.9 & \cellcolor[gray]{.8} 5.5 &  5.1 &  5.2 &  5.0 &  4.8&\\
  6400 &  5.3 &  5.0 &  5.1 &  4.8 & \cellcolor[gray]{.8} 5.6 &  5.1 &  5.4 &  5.3 &  5.3 &  5.0 &  5.4 &  4.9&\\
   \hline
   200 &  5.2 &  5.1 &  5.1 &  5.0 &  5.2 &  4.8&  5.2 &  4.9 &  5.1 &  5.2 &  5.3 &  5.1& \multirow{3}{*}{$\frac{2p(x)}{g_2(x)}$}\\
   400 &  5.1 &  5.3 &  5.4 &  5.0 &  4.9 &  4.9 &  5.3 &  5.3 & \cellcolor[gray]{.8} 5.5 &  5.1 &  4.9 &  5.0&\\
   800 &  5.0 &  4.9 &  4.5 &  4.7 &  5.2 &  5.0 &  5.1 &  4.8 &  4.9 & \cellcolor[gray]{.8} 4.5 &  4.8 &  5.2&\multirow{2}{*}{$\&$}\\
  1600 &  4.9 &  5.0 &  4.9 &  5.1 &  4.9 &  4.6 &  5.3 &  5.2 &  5.2 &  5.3 &  4.9 &  4.6& \multirow{3}{*}{$\frac{3p(x)}{g_3(x)}$}\\
  3200 &  4.9 &  5.1 &  5.3 &  5.0 &  4.8 &  4.8 &  4.8 &  5.2 &  5.2 &  5.2 &  4.7 &  5.2&\\
  6400 &  5.1 &  5.0 &  4.9 &  5.2 &  4.8 & \cellcolor[gray]{.8} 4.5 &  5.3 &  4.9 &  4.7 &  5.0 &  5.0 &  4.8&\\
   \hline
   200 &  5.0 &  4.8 &  4.9 &  5.0 &  4.7 &  4.8 &  5.1 &  4.7 &  5.0 &  5.1 & \cellcolor[gray]{.8} 5.6 &  5.2& \multirow{3}{*}{$\frac{2p(x)}{g_3(x)}$}\\
   400 &  5.3 &  5.1 &  4.8 & \cellcolor[gray]{.8} 5.5 &  4.9 &  4.8 &  5.3 &  4.9 &  4.9 &  5.0 &  5.0 &  5.1&\\
   800 &  5.2 &  5.3 &  4.6 &  4.9 &  5.1 &  5.1 &  5.1 &  4.8 &  5.3 &  5.2 &  4.9 &  4.9&\multirow{2}{*}{$\&$}\\
  1600 &  5.3 &  5.0 &  4.8 &  5.0 &  5.0 &  5.1 &  4.9 &  5.0 &  5.3 &  4.9 &  4.7 &  4.8& \multirow{3}{*}{$\frac{3p(x)}{g_3(x)}$}\\
  3200 &  4.9 & \cellcolor[gray]{.8} 5.5 &  4.9 &  4.9 &  5.0 &  5.2 &  5.0 &  4.8 &  4.8 &  5.3 &  5.4 &  5.0&\\
  6400 &  5.2 &  5.0 &  5.1 &  5.2 &  5.4 &  5.1 & \cellcolor[gray]{.8} 5.5 &  5.2 &  5.0 &  4.9 &  4.9 &  4.7&\\
\hline
\end{tabular}
\end{table}
\vspace *{-1.4 cm}
\begin{table}[H]
\centering
\caption{Sizes $\hat\alpha_s$  of the  test $X^2_{p2}$ (\ref{pois})  for comparison of two Poisson weighted  histograms with normalized and  unnormalized weighted entries (left panel) and sizes of the new test  $_1\hat {\check{X}}^2$(\ref{stddc5})  (right panel)  for different pairs of weights (last column)  and parameters  $n_{01}$, $n_{02}$.  Sizes of the tests that do not satisfy the hypothesis    $\alpha_s=5\%$     with a significance level equal to $0.05$   ($\hat\alpha_s>5.4\%$ or $\hat\alpha_s<4.6\%$) are highlighted with gray.}
\label{tab:result3c}
\vspace *{0.2 cm}
\tiny
\begin{tabular}{r|rrrrrr||rrrrrr|c}
& \multicolumn{6}{c||}{$n_{02}$} & \multicolumn{6}{c|}{$n_{02}$} & \\
$n_{01}$&  200 &  400 &  800 & 1600 & 3200 & 6400 &  200 &  400 &  800 & 1600 & 3200 & 6400&$w(x)$\\
\hline
  200 &  5.1 & \cellcolor[gray]{.8} 4.5 &  4.8 & \cellcolor[gray]{.8} 4.5 &  5.2 & \cellcolor[gray]{.8} 4.4 &  5.0 &  5.1 &  4.7 &  4.9 &  5.3 & \cellcolor[gray]{.8} 4.5& \multirow{3}{*}{$1$}\\
   400 &  5.2 &  4.8 &  5.1 &  5.3 &  4.7 &  4.7 &  5.1 &  5.1 &  5.2 &  5.1 & \cellcolor[gray]{.8} 4.5 &  5.4&\\
   800 &  5.1 &  4.9 &  5.4 &  4.6 &  4.9 &  5.1 &  5.1 &  5.2 &  5.1 &  4.6 &  4.7 &  4.8&\multirow{2}{*}{$\&$}\\
  1600 &  4.8 &  4.7 &  5.0 &  4.8 &  4.9 & \cellcolor[gray]{.8} 4.5 &  5.1 &  5.2 &  5.0 &  5.1 &  4.9 &  5.0& \multirow{3}{*}{$\frac{3p(x)}{g_2(x)}$}\\
  3200 &  4.9 &  4.8 &  5.1 &  4.9 &  4.9 &  4.9 &  5.4 &  4.8 &  4.9 &  4.8 &  5.0 &  4.9&\\
  6400 &  5.2 &  5.1 &  4.9 &  5.0 &  4.8 &  5.0 &  5.0 &  5.1 &  4.9 &  5.1 &  4.7 &  4.7&\\
  \hline
   200 &  5.1 &  4.9 &  4.8 &  4.6 &  4.8 &  4.8 &  4.9 &  4.8 &  4.8 &  4.9 &  4.7 & \cellcolor[gray]{.8} 5.6& \multirow{3}{*}{$1$}\\
   400 &  4.7 &  4.8 & \cellcolor[gray]{.8} 4.4 &  5.2 &  4.7 &  4.9 &  4.8 &  4.6 &  4.9 & \cellcolor[gray]{.8} 4.5 &  5.3 &  5.1&\\
   800 &  5.2 &  4.9 &  5.2 &  4.9 &  5.2 &  5.2 &  5.3 &  5.0 &  5.4 &  4.8 &  5.2 &  4.8&\multirow{2}{*}{$\&$}\\
  1600 &  4.9 &  5.2 &  5.0 &  4.7 &  5.2 &  5.0 &  5.2 &  5.1 &  5.0 &  5.2 &  5.0 &  5.0& \multirow{3}{*}{$\frac{3p(x)}{g_3(x)}$}\\
  3200 &  5.2 &  4.8 &  4.6 &  4.9 &  5.3 &  4.6 &  5.2 &  4.7 &  4.9 &  5.0 &  4.6 &  4.6&\\
  6400 &  5.0 &  5.1 &  5.0 &  5.1 &  4.8 &  5.0 &  5.0 &  4.9 &  4.9 &  5.1 &  4.9 &  4.7&\\
  \hline
   200 &  5.2 &  5.1 & \cellcolor[gray]{.8} 5.5 &  4.7 &  5.2 &  5.2 &  4.8 &  4.7 &  5.1 &  5.3 &  5.0 &  5.4& \multirow{3}{*}{$\frac{p(x)}{g_2(x)}$}\\
   400 &  4.9 &  5.4 &  4.9 &  5.2 &  4.9 &  5.3 &  5.2 &  5.1 &  4.8 &  5.2 &  4.9 &  4.7&\\
   800 &  5.2 &  5.2 &  4.9 &  5.3 &  5.1 &  5.1 &  4.7 &  5.3 &  5.4 &  5.2 &  4.5 &  5.3&\multirow{2}{*}{$\&$}\\
  1600 &\cellcolor[gray]{.8}  4.5 &  5.3 &  5.1 &  4.9 &  5.2 &  5.0 &  5.4 &  4.9 &  5.2 &  5.3 &  4.7 &  5.1& \multirow{3}{*}{$\frac{3p(x)}{g_2(x)}$}\\
  3200 &  5.1 &  4.9 &  4.9 &  4.7 &  4.9 &  4.7&  5.4 & \cellcolor[gray]{.8} 5.3 & \cellcolor[gray]{.8} 4.5 &  5.0 &  5.0 &  5.2&\\
  6400 &  4.7 &  5.2 &  5.1 &  5.1 &  5.4 &  5.4 &  4.9 &  5.4 &  5.2 &  4.8 &  4.9 &  5.1&\\
   \hline
   200 &  5.3 &  4.7 &  5.2 &  5.1 &  5.1 &  4.8 &  5.5 &  5.7 &  5.4 &  5.2 &  5.0 & \cellcolor[gray]{.8} 5.7& \multirow{3}{*}{$\frac{p(x)}{g_2(x)}$}\\
   400 &  5.0 &  4.9 &  4.9 & \cellcolor[gray]{.8} 5.6 &  5.0 &\cellcolor[gray]{.8}  5.5 &  5.1 &  5.4 &  5.3 &  4.9 &  5.0 &  5.1&\\
   800 &  4.9 &  5.2 &  4.8 &  5.0 &  4.6 &  5.3 &  5.2 & \cellcolor[gray]{.8} 5.5 &  5.1 &  4.8 &  4.6 &  4.9&\multirow{2}{*}{$\&$}\\
  1600 &  5.1 &  4.7 &  4.7 &  5.2 &  5.2 &  4.7 &  5.4 &  4.8 &  5.1 &  4.9 &  5.4 &  4.8& \multirow{3}{*}{$\frac{3p(x)}{g_3(x)}$}\\
  3200 &  4.6 &  4.8 & \cellcolor[gray]{.8} 5.8 &  5.1 &  5.1 &  5.1 &  5.0 &  5.3 &  5.1 &  5.2 &  4.9 &  4.9&\\
  6400 &  5.3 &  4.9 &  4.9 &  5.2 &  4.9 &  5.2 &  4.8 &  5.4 &  5.0 &  5.0 &  5.2 &  4.9&\\
   \hline
   200 &  5.1 &  5.1 &  4.7 &  4.8 &  5.0 &  4.9 &  5.4 & \cellcolor[gray]{.8} 5.9 &  5.1 &  5.4 &  5.4 &  5.4& \multirow{3}{*}{$\frac{p(x)}{g_3(x)}$}\\
   400 &  5.0 &  5.2 &  4.9 &  5.3 &  4.6 &  4.9 &  5.0 &  5.2 &  5.4 &  4.8 &  5.0 &  4.9&\\
   800 &  5.2 &  4.8 &  4.8 &  5.1 &  5.3 &  5.1 &  5.1 &  5.3 &  5.0 &  5.4 &  5.1 &  5.3&\multirow{2}{*}{$\&$}\\
  1600 &  4.6 &  4.9 &  4.8 &  5.3 &  4.9 & \cellcolor[gray]{.8} 4.5 &  4.7 &  4.8 &  4.9 &  4.6 &  5.2 &  5.0& \multirow{3}{*}{$\frac{3p(x)}{g_2(x)}$}\\
  3200 &  5.1 &  5.1 &  5.0 &  5.4 &  4.8 &  4.9 &  5.4 &  4.7 &  5.3 &  5.2 &  5.3 &  4.8&\\
  6400 &  4.8 &  4.9 &  4.9 &  5.1 &  5.3 &  5.1&  5.2 &  5.3 & \cellcolor[gray]{.8} 5.6 &  5.0 &  4.7 &  4.6&\\
   \hline
   200 &  5.1 &  4.9 &  4.9 &  4.8 &  5.1 &  4.9 &  5.3 &  5.4 & \cellcolor[gray]{.8} 5.7 & \cellcolor[gray]{.8} 6.0 & \cellcolor[gray]{.8} 5.6 &  5.0& \multirow{3}{*}{$\frac{p(x)}{g_3(x)}$}\\
   400 &  5.0 &  4.9 &  5.0 &  4.8 &  4.7 &  4.9 &  5.1 &  5.3 &  5.0 & \cellcolor[gray]{.8} 5.5 &  5.3 &  4.7&\\
   800 &  4.7 &  4.9 &  5.1 &  5.2 &  5.3 &  4.9 &  5.4 &  5.1 &  5.0 &  5.4 &  4.8 &  4.9&\multirow{2}{*}{$\&$}\\
  1600 &  4.6 &  5.2 &  5.3 &  5.0 &  5.2 &  5.0 &  5.0 &  5.1 & \cellcolor[gray]{.8} 4.5 &  4.7 &  4.9 &  5.1& \multirow{3}{*}{$\frac{3p(x)}{g_3(x)}$}\\
  3200 &  5.0 &  4.8 &  4.7 &  4.8 &  5.1 &  5.0 &  5.1 &  4.9 & \cellcolor[gray]{.8} 5.5 &  5.1 &  5.0 &  5.1&\\
  6400 &  5.0 &  4.9 &  4.9 &  5.0 &  4.9 &  4.8 &  5.4 &  4.8 &  5.1 &  5.2 &  4.9 &  5.1&\\
\hline
\end{tabular}
\end{table}
Distribution  of p-value was studied by simulating 100\,000 runs.  In each run  a number of events was simulated according to a Poisson distribution with the parameter $n_{01} =3\,200$ for the first histogram  and $n_{02}= 6\,400$ for the second one.  The first  histogram  represents  the $p(x)$  distribution  with weights  $\frac{2p(x)}{g_2(x)}$  and  the second histogram  represents   the  $p(x)$  distribution of event   with weights $\frac{3p(x)}{g_3(x)}$. To compare two Poisson weighted  histograms with unnormalized weights the  new statistic           $\hat {\check{X}}^2$ (\ref{stddc5}),  the first statistic $X^2_{p1}$ (\ref{pois})  \cite{zex}   and the  second statistic  $X^2_{p2}$ (\ref{pois2}) \cite{zex}  were used. 
\newpage
\vspace *{-2 cm}
\begin{figure}[H]
\includegraphics[width=1.05 \textwidth]{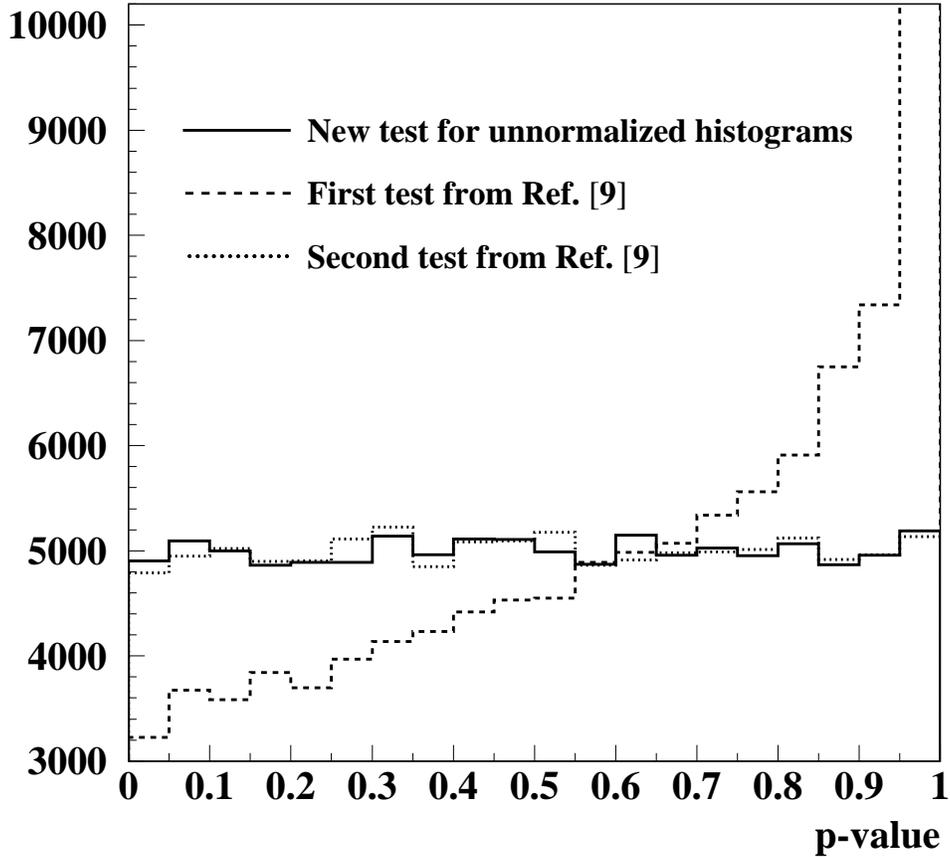}\\
\vspace*{-1.5 cm}
\caption {Distributions of p-value for the new statistic   $\hat {\check{X}}^2$ (\ref{stddc5}), the first statistic $X^2_{p1}$ (\ref{pois})  \cite{zex}    and  the second statistic  $X^2_{p2}$ (\ref{pois2}) \cite{zex} used for comparison of two Poisson  weighted histograms with unnormalized weights.}
\vspace*{1 cm }
\end{figure}

\noindent
{\it Conclusions to subsection 6.2.1}
\begin{itemize}
\item Tables 4-6\\
 The sizes $\hat\alpha_s$  of the new tests $\hat X^2$ (\ref{stdd3}), $\hat {\check{X}}^2$, $_1\hat {\check{X}}^2$ (\ref{stddc5}) for comparison of Poisson weighted histograms  are close to a nominal value of a test size equal to 5\%  as well as the  sizes of test  $X^2_{p2}$ (\ref{pois2}) \cite{zex}.
\newpage
\item Figure 5\\
The distribution of the new statistic $\hat {\check{X}}^2$ (\ref{stddc5})  for comparison of weighted histograms with unnormalized weights is  close to  a $\chi^2_{m-2}$ distribution while  the distribution of the statistic $X^2_{p2}$ (\ref{pois2}) \cite{zex} is close to  a $\chi^2_{m-1}$ distribution.

Assumption that the statistic   $X^2_{p1}$ (\ref{pois})  \cite{zex} has  a $\chi^2_{m}$   distribution is wrong and  the  statistic    $X^2_{p1}$ (\ref{pois})  \cite{zex} cannot be recommended for use   in data analysis.
\end{itemize} 
\subsubsection {Power of tests for comparison of Poisson weighted histograms }
 Calculation of power was performed for the specified probability distribution function $p_0(x)$  (\ref{weight3}).
\begin{table}[H]
\centering
\caption{Power $\beta$ of the new test  $\hat X^2$ (\ref{stdd3}) used for  comparison of two Poisson histograms with normalized weighted entries (right panel) and the exceedance of power of the test  $\hat X^2$(\ref{stdd3})  over the power of the  test  $X^2_{p2}$ (\ref{pois2}) (left panel) for different pairs of weights (last column)  and  parameters  $n_{01}$, $n_{02}$.  
Cases  when the power of the test $X^2_{p2}$ (\ref{pois2}) exceeds  the power of the new test $\hat X^2$ (\ref{stdd3}) are highlighted with gray.}
\label{tab:result1pw}
\vspace *{0.2 cm}
\tiny
\begin{tabular}{r|rrrrrr||rrrrrr|c}
& \multicolumn{6}{c||}{$n_{02}$} & \multicolumn{6}{c|}{$n_{02}$} & \\
$n_{01}$&  200 &  400 &  800 & 1600 & 3200 & 6400 &  200 &  400 &  800 & 1600 & 3200 & 6400&$w(x)$\\
\hline
 200 & \cellcolor[gray]{.8} -0.4 & \cellcolor[gray]{.8} -0.1 & \cellcolor[gray]{.8} -0.1 &  0.2 &  0.4 &  0.8    &  6.0 &  6.1 &  6.8 &  7.2 &  7.2 &  7.6& \multirow{3}{*}{$1$}\\
   400 & \cellcolor[gray]{.8}-0.2&	0.2&	0.3&	0.2&	0.3&	\cellcolor[gray]{.8}-0.1 &  5.8 &  6.7 &  8.4 &  8.7 &  9.3 &  9.1&\\
   800 &\cellcolor[gray]{.8} -0.1&\cellcolor[gray]{.8}	-0.7&	\cellcolor[gray]{.8}-0.2&	0.6&	0.2&	0.2  &  6.6 &  7.5 &  9.7 & 11.5 & 13.9 & 14.8&\multirow{2}{*}{$\&$}\\
  1600 & 0.4&	\cellcolor[gray]{.8}-0.2&	\cellcolor[gray]{.8}-0.2&\cellcolor[gray]{.8}	-0.1&	0.0&	\cellcolor[gray]{.8}-0.1  &  6.9 &  8.4 & 11.4 & 15.8 & 20.0 & 24.2& \multirow{3}{*}{$1$}\\
  3200 & 0.1&	0.1&	0.0&	\cellcolor[gray]{.8}-0.1&	0.0&	\cellcolor[gray]{.8}-0.1  &  7.3 & 10.2 & 12.9 & 20.3 & 28.1 & 38.7&\\
  6400 & 0.2&	0.1&	0.1&	0.0&	0.0&	0.0 &  6.8 &  9.6 & 14.0 & 24.0 & 37.8 & 56.1&\\
  \hline
   200 &\cellcolor[gray]{.8}-0.3&	\cellcolor[gray]{.8}-0.2&\cellcolor[gray]{.8}	-0.1&	0.1&	0.0&	0.3 &  6.2 &  6.6 &  6.2 &  6.5 &  6.6 &  6.9& \multirow{3}{*}{$1$}\\
   400 &\cellcolor[gray]{.8} -0.4&	0.2&	0.0&	0.1&	0.7&	0.2  &  6.3 &  7.4 &  7.7 &  8.9 &  9.3 &  9.5&\\
   800 &\cellcolor[gray]{.8} -0.7&	0.1&	\cellcolor[gray]{.8}-0.7&	0.0&	\cellcolor[gray]{.8}-0.6&\cellcolor[gray]{.8}	-0.7 &  6.4 &  8.0 &  8.5 & 11.1 & 11.8 & 13.6&\multirow{2}{*}{$\&$}\\
  1600 & 0.2&\cellcolor[gray]{.8}	-0.3&	0.1&	0.5&	0.7&	\cellcolor[gray]{.8}-0.1  &  7.2 &  8.2 & 10.8 & 14.2 & 18.2 & 22.8& \multirow{3}{*}{$\frac{p_0(x)}{g_2(x)}$}\\
  3200 &0.4&	0.2&	0.0&\cellcolor[gray]{.8}	-0.8&	0.1&	0.9 &  7.3 &  8.8 & 11.2 & 16.5 & 25.3 & 34.7&\\
  6400 & 1.0&	0.6&	0.1&	\cellcolor[gray]{.8}	-0.8&	0.3&	0.5  &  7.5 &  9.0 & 11.5 & 18.7 & 32.0 & 47.9&\\
  \hline
   200 & \cellcolor[gray]{.8}-0.4&\cellcolor[gray]{.8}	-0.8&\cellcolor[gray]{.8}	-0.2&	0.1&	0.3&	0.0  &  5.7 &  6.0 &  6.5 &  6.6 &  6.9 &  6.9& \multirow{3}{*}{$1$}\\
   400 &0.2&	0.4&	0.1&	0.5&	\cellcolor[gray]{.8}-0.3&	0.0 &  6.4 &  7.2 &  8.2 &  9.1 &  8.8 &  9.1&\\
   800 & 0.4&\cellcolor[gray]{.8}	-0.4&	0.5&\cellcolor[gray]{.8}	-0.5&\cellcolor[gray]{.8}	-0.2&	0.9 &  7.2 &  7.8 &  9.8 & 11.1 & 13.2 & 14.6&\multirow{2}{*}{$\&$}\\
  1600 & 0.5&	0.0&	0.3&	0.2&	0.9&	0.2  &  6.9 &  8.7 & 11.5 & 15.2 & 19.9 & 23.4& \multirow{3}{*}{$\frac{p_0(x)}{g_3(x)}$}\\
  3200 & 0.9&	0.3&	\cellcolor[gray]{.8}-0.3&	0.7&	0.9&	0.8 &  7.1 &  9.4 & 13.0 & 19.6 & 27.8 & 36.6&\\
  6400 & 0.6&	\cellcolor[gray]{.8}-0.1&	\cellcolor[gray]{.8}-0.2&	0.8&	2.3&	2.1  &  7.5 &  9.0 & 13.6 & 23.1 & 37.5 & 53.7&\\
   \hline
   200 & 0.5&	0.1&	0.0&	0.3&	0.8&	0.6  &  6.4 &  6.5 &  6.7 &  7.1 &  7.7 &  7.7& \multirow{3}{*}{$\frac{p(x)}{g_2(x)}$}\\
   400 & 0.3&	0.2&	0.5&	0.1&\cellcolor[gray]{.8}	-0.6&\cellcolor[gray]{.8}	-0.1 &  6.4 &  7.2 &  7.7 &  8.2 &  8.1 &  8.7&\\
   800 & 0.3&	\cellcolor[gray]{.8}-0.5&	\cellcolor[gray]{.8}-0.3&	0.1	&0.1 &	0.1  &  6.7 &  6.8 &  8.6 & 10.0 & 11.7 & 12.0&\multirow{2}{*}{$\&$}\\
  1600 & 0.1&\cellcolor[gray]{.8}	-0.1&	\cellcolor[gray]{.8}-1.0&	0.2&	0.7&	\cellcolor[gray]{.8}-0.7 &  6.9 &  7.9 &  9.8 & 13.2 & 17.0 & 18.9& \multirow{3}{*}{$\frac{p_0(x)}{g_2(x)}$}\\
  3200 & 0.3&	0.9&	0.3&	\cellcolor[gray]{.8}-0.4&	\cellcolor[gray]{.8}-0.3&\cellcolor[gray]{.8}	-0.5 &  6.9 &  8.6 & 11.4 & 16.4 & 23.1 & 29.7&\\
  6400 & 0.8 &\cellcolor[gray]{.8}-0.5&	\cellcolor[gray]{.8}-0.6&	\cellcolor[gray]{.8}-0.2&	\cellcolor[gray]{.8}-0.4&	\cellcolor[gray]{.8}-1.9 &  7.7 &  7.7 & 12.3 & 18.2 & 29.8 & 44.5&\\
   \hline
   200 & 0.2 &	0.4&	0.4&	0.3 &	0.3	&0.7 &  6.3 &  6.8 &  7.2 &  6.9 &  7.4 &  7.3& \multirow{3}{*}{$\frac{p(x)}{g_2(x)}$}\\
   400 & 0.4 &	0.1 &	0.7 &	0.5 &	0.2 &	0.1 &  6.4 &  7.2 &  8.0 &  8.5 &  8.8 &  8.9&\\
   800 & 1.0 &\cellcolor[gray]{.8}	-0.5 &	0.3	 & 0.0&	0.2&\cellcolor[gray]{.8}	-0.1 &  7.2 &  7.3 &  9.1 & 10.4 & 11.9 & 11.7&\multirow{2}{*}{$\&$}\\
  1600 & 0.8 &	0.3 &	1.6 &	0.1	&0.3	&0.5  &  6.9 &  8.5 & 11.5 & 13.6 & 16.7 & 18.7& \multirow{3}{*}{$\frac{p_0(x)}{g_3(x)}$}\\
  3200 & \cellcolor[gray]{.8}-0.2 &0.2&0.3 &1.4 &	0.8 &	1.2 &  6.9 &  8.8 & 12.3 & 18.8 & 25.1 & 31.3&\\
  6400 & \cellcolor[gray]{.8}-0.2&0.3&0.7&2.2 &	1.0 &	0.8  &  7.1 &  9.6 & 14.0 & 22.1 & 34.4 & 48.3&\\
  \hline
   200 & 0.3 &	0.9	&1.4&	2.1 &	2.3 &	2.2 &  6.7 &  7.5 &  8.3 &  9.5 &  9.7 &  9.7& \multirow{3}{*}{$\frac{p(x)}{g_3(x)}$}\\
   400 & 0.6 &	1.0 &	0.8 &	0.9	&1.8 &	2.2 &  6.7 &  7.9 &  9.1 & 10.3 & 10.7 & 11.5&\\
   800 & 0.5&	0.0&	0.9&	0.0&	0.9 &	0.8 &  7.1 &  8.3 & 10.3 & 12.5 & 14.0 & 16.0&\multirow{2}{*}{$\&$}\\
  1600 & 0.5 &\cellcolor[gray]{.8}	-0.3&	0.2&	0.6&	2.6	&3.3  &  7.1 &  8.4 & 11.2 & 15.7 & 21.1 & 25.9& \multirow{3}{*}{$\frac{p_0(x)}{g_3(x)}$}\\
  3200 & 0.6&	0.0	&\cellcolor[gray]{.8}-0.1	&1.7 &	2.4	 & 3.4  &  7.5 &  9.1 & 13.2 & 20.6 & 30.6 & 40.4&\\
  6400 & 0.8 &	0.7	&0.3	&0.5&	1.7&	2.4  &  7.4 &  9.5 & 13.7 & 23.2 & 38.1 & 55.2&\\
\hline
\end{tabular}
\end{table}
\vspace *{-4.cm}
\begin{table}[H]
\centering
\caption{ Power $\beta$ of  the new test  $\hat {\check{X}}^2$ (\ref{stddc5})   used  for comparison of two Poisson histograms with unnormalized weighted entries (right panel) and the exceedance of power of  the test   $\hat {\check{X}}^2$ (\ref{stddc5}) over the power of the test  $X^2_{p2}$ (\ref{pois2}) (left panel) for different pairs of weights (last column)  and  parameters  $n_{01}$, $n_{02}$.  Cases  when the power of the test $X^2_{p2}$ (\ref{pois2}) exceeds  the power of the new test $\hat {\check{X}}^2$ (\ref{stddc5})  are highlighted with gray.}
\label{tab:result2pw}
\vspace *{.2cm}
\tiny
\begin{tabular}{r|rrrrrr||rrrrrr|c}
& \multicolumn{6}{c||}{$n_{02}$} & \multicolumn{6}{c|}{$n_{02}$} & \\
$n_{01}$&  200 &  400 &  800 & 1600 & 3200 & 6400 &  200 &  400 &  800 & 1600 & 3200 & 6400&$w(x)$\\
\hline
 200 &\cellcolor[gray]{.8}-0.3&0.0&0.0&0.1&0.4&0.4  &  5.7 &  6.4 &  6.5 &  7.0 &  7.0 &  7.3& \multirow{3}{*}{$\frac{2p(x)}{g_2(x)}$}\\
   400 & 0.2&\cellcolor[gray]{.8}-0.2&0.1&0.4&0.5&0.4  &  6.0 &  6.8 &  7.9 &  8.2 &  9.4 &  8.9&\\
   800 &0.3&0.5&0.2&0.2&0.5&0.4  &  6.7 &  7.6 &  9.2 & 11.0 & 11.4 & 13.2&\multirow{2}{*}{$\&$}\\
  1600 & 0.2&0.5&\cellcolor[gray]{.8}-0.2&0.4&0.5&1.1&  6.5 &  8.4 & 10.6 & 14.0 & 16.2 & 20.5& \multirow{3}{*}{$\frac{3p_0(x)}{g_2(x)}$}\\
  3200 &0.2&0.1&0.6&0.9&0.5&0.7  &  7.1 &  8.9 & 12.0 & 17.9 & 23.6 & 32.2&\\
  6400 & 0.4&0.3&0.5&0.9&0.4&0.3 &  7.4 &  8.9 & 12.3 & 20.7 & 30.8 & 45.3&\\
 \hline
   200 & 0.3&\cellcolor[gray]{.8}-0.2&0.3&0.4&0.5&0.5 &  6.4 &  6.0 &  6.7 &  7.0 &  7.5 &  7.1& \multirow{3}{*}{$\frac{2p(x)}{g_2(x)}$}\\
   400 &0.4&0.2&0.6&0.3&0.3&0.5 &  6.6 &  7.4 &  8.3 &  8.3 &  8.7 &  9.0&\\
   800 & 0.7&0.5&0.8&0.5&0.6&1.0 &  7.1 &  8.2 &  9.5 & 10.6 & 12.2 & 13.7&\multirow{2}{*}{$\&$}\\
  1600 & 0.8&1.0&1.2&1.3&1.5&1.1 &  7.2 &  9.4 & 11.6 & 15.1 & 18.5 & 20.7& \multirow{3}{*}{$\frac{3p_0(x)}{g_3(x)}$}\\
  3200 & 0.8&0.7&1.5&2.0&2.5&2.8  &  7.2 &  9.9 & 13.9 & 20.1 & 25.8 & 34.0&\\
  6400 & 0.9&0.9&1.2&2.5&3.3&3.6 &  7.7 & 10.1 & 14.1 & 24.0 & 37.2 & 50.8&\\
  \hline
   200 & \cellcolor[gray]{.8} -0.3&0.1&0.7&0.1&1.1&0.7 &  6.2 &  6.3 &  7.5 &  7.8 &  7.9 &  8.2& \multirow{3}{*}{$\frac{2p(x)}{g_3(x)}$}\\
   400 & 0.5&0.7&0.0&0.5&0.5&0.6 &  6.8 &  7.9 &  8.0 &  9.9 & 10.2 & 10.0&\\
   800 & 0.2&0.6&1.2&1.2&1.8&1.5 &  7.4 &  8.6 & 10.9 & 12.4 & 15.0 & 15.2&\multirow{2}{*}{$\&$}\\
  1600 & 0.2&0.5&1.5&1.4&1.5&1.8 &  7.1 &  9.5 & 12.8 & 16.3 & 21.1 & 25.8& \multirow{3}{*}{$\frac{3p_0(x)}{g_3(x)}$}\\
  3200 & 1.1&0.3&1.2&2.3&3.7&2.9  &  7.3 &  9.9 & 13.9 & 21.2 & 30.8 & 40.0&\\
  6400 & 1.1&1.6&2.2&2.0&3.5&3.8  &  8.4 & 10.6 & 16.1 & 24.6 & 40.3 & 56.6&\\
  \hline
\end{tabular}
\end{table}
\vspace *{0.7cm}
\begin{table}[H]
\centering
\caption{ Power $\beta$ of  the new test $_1\hat {\check{X}}^2$ (\ref{stddc5}) used for comparison of two Poisson histograms with normalized and unnormalized weighted  entries (right panel)   and the exceedance of the power of the test $_1\hat {\check{X}}^2$ (\ref{stddc5})   over the power of the  test  $X^2_{p2}$ (\ref{pois2}) (left panel) for different pairs of weights (last column)  and     parameters $n_{01}$, $n_{02}$. Cases  when the power of the test $X^2_{p2}$ (\ref{pois2}) exceeds the  power of the new test $_1\hat {\check{X}}^2$ (\ref{stddc5})  are highlighted with gray. }
\label{tab:result3pw}
\vspace *{0.2cm}
\scriptsize
\tiny
\begin{tabular}{r|rrrrrr||rrrrrr|c}
& \multicolumn{6}{c||}{$n_{02}$} & \multicolumn{6}{c|}{$n_{02}$} & \\
$n_{01}$&  200 &  400 &  800 & 1600 & 3200 & 6400 &  200 &  400 &  800 & 1600 & 3200 & 6400&$w(x)$\\
\hline
 200 & 0.0&0.4&0.1&0.2&0.4&0.3&  6.1 &  6.6 &  6.8 &  6.8 &  7.4 &  6.7& \multirow{3}{*}{$1$}\\
   400 & 0.3&0.4&0.4&\cellcolor[gray]{.8}-0.2&0.7&0.5  &  6.8 &  7.3 &  8.8 &  8.3 &  9.5 &  9.8&\\
   800 & 0.1&0.2&\cellcolor[gray]{.8}-0.2&0.7&1.1&1.4  &  7.2 &  7.8 &  9.7 & 11.5 & 13.8 & 15.3&\multirow{2}{*}{$\&$}\\
  1600 & 0.3&0.2&0.9&0.9&1.7&3.0  &  6.8 &  8.4 & 11.5 & 15.2 & 19.6 & 25.1& \multirow{3}{*}{$\frac{3p_0(x)}{g_2(x)}$}\\
  3200 &0.4&0.7&0.3&1.4&1.7&3.2 &  7.5 &  9.0 & 12.1 & 18.1 & 25.4 & 36.5&\\
  6400 & 0.6&0.3&1.0&1.0&1.7&3.7 &  7.5 &  8.8 & 12.9 & 20.5 & 32.5 & 50.2&\\
  \hline
   200 &  0.1&\cellcolor[gray]{.8}-0.1&0.3&0.4&0.5&0.6 &  6.2 &  6.7 &  6.8 &  7.3 &  7.3 &  7.3& \multirow{3}{*}{$1$}\\
   400 & 1.0&0.2&0.5&0.3&1.0&0.3  &  7.2 &  7.2 &  7.9 &  9.2 & 10.2 &  9.6&\\
   800 & 0.8&0.5&0.5&0.9&1.2&1.6  &  7.5 &  8.6 & 10.5 & 12.4 & 14.2 & 15.7&\multirow{2}{*}{$\&$}\\
  1600 & 0.1&0.8&0.9&1.9&2.0&2.3  &  7.1 &  9.4 & 11.8 & 16.6 & 21.6 & 25.8& \multirow{3}{*}{$\frac{3p_0(x)}{g_3(x)}$}\\
  3200 & 0.7&0.7&1.9&1.7&3.0&2.7  &  7.7 &  9.9 & 14.1 & 20.6 & 30.2 & 40.6&\\
  6400 & 0.2&0.5&1.5&3.0&3.5&4.3 &  7.3 & 10.2 & 14.2 & 25.9 & 39.5 & 56.6&\\
  \hline
   200 &\cellcolor[gray]{.8} -0.1&0.2&0.0&0.4&0.3&0.2  &  5.8 &  6.4 &  6.8 &  6.5 &  6.6 &  7.1& \multirow{3}{*}{$\frac{p(x)}{g_2(x)}$}\\
   400 &0.4&\cellcolor[gray]{.8}-0.1&\cellcolor[gray]{.8}-0.2&\cellcolor[gray]{.8}-0.5&0.4&0.1 &  6.7 &  6.8 &  7.4 &  7.8 &  8.6 &  8.5&\\
   800 & 0.4&\cellcolor[gray]{.8}-0.3&0.3&0.5&0.7&0.1 &  7.2 &  7.8 &  9.1 & 10.3 & 12.0 & 12.5&\multirow{2}{*}{$\&$}\\
  1600 & 0.6&0.5&\cellcolor[gray]{.8}-0.3&1.1&0.3&0.6 &  6.8 &  9.0 & 10.1 & 13.9 & 17.2 & 20.4& \multirow{3}{*}{$\frac{3p_0(x)}{g_2(x)}$}\\
  3200 & \cellcolor[gray]{.8}-0.5&0.4&\cellcolor[gray]{.8}-0.3&1.2&1.7&\cellcolor[gray]{.8}-0.6&  6.3 &  9.1 & 11.0 & 16.9 & 24.6 & 30.7&\\
  6400 & 0.9&0.6&0.7&1.7&0.9&2.0  &  7.3 &  9.3 & 13.0 & 20.8 & 31.5 & 45.7&\\
  \hline
   200 & 0.4&0.8&1.1&0.8&0.9&1.2 &  6.7 &  7.2 &  7.6 &  7.5 &  7.8 &  7.8& \multirow{3}{*}{$\frac{p(x)}{g_2(x)}$}\\
   400 & 0.3&1.1&0.5&0.2&0.7&0.2  &  6.5 &  7.6 &  8.0 &  8.9 &  8.6 &  9.2&\\
   800 &  0.8&0.4&0.2&0.3&0.8&1.1 &  7.0 &  7.9 &  9.6 & 10.8 & 12.2 & 13.0&\multirow{2}{*}{$\&$}\\
  1600 &1.1&1.5&0.6&0.8&0.9&1.7 &  7.6 &  9.3 & 11.5 & 14.5 & 17.8 & 21.1& \multirow{3}{*}{$\frac{3p_0(x)}{g_3(x)}$}\\
  3200 & 0.9&1.3&1.3&1.8&2.5&2.8  &  7.2 &  9.9 & 14.0 & 19.6 & 27.5 & 33.9&\\
  6400 & 0.6&1.0&2.0&2.5&3.7&3.3  &  7.5 &  9.6 & 16.0 & 23.8 & 36.7 & 51.1&\\
  \hline
   200 & 0.8&0.7&1.5&1.1&1.7&1.4  &  7.0 &  7.3 &  8.0 &  8.2 &  9.2 &  8.7& \multirow{3}{*}{$\frac{p(x)}{g_3(x)}$}\\
   400 &0.4&0.7&0.7&0.1&1.6&1.0  &  6.8 &  7.9 &  8.6 &  9.4 & 10.4 & 10.1&\\
   800 & \cellcolor[gray]{.8}-0.4&0.5&0.4&0.7&1.6&1.2  &  6.8 &  8.1 &  9.6 & 11.9 & 14.1 & 14.6&\multirow{2}{*}{$\&$}\\
  1600 & 0.8&\cellcolor[gray]{.8}-0.6&0.9&0.8&1.9&2.7 &  7.2 &  7.8 & 10.9 & 14.6 & 19.1 & 23.1& \multirow{3}{*}{$\frac{3p_0(x)}{g_2(x)}$}\\
  3200 &\cellcolor[gray]{.8} -0.2&0.3&0.3&1.5&1.8&2.9 &  6.9 &  8.7 & 11.6 & 18.9 & 27.2 & 36.2&\\
  6400 & 0.7&0.6&1.0&0.5&1.8&2.6  &  7.4 &  9.4 & 13.8 & 20.6 & 33.7 & 50.1&\\
  \hline
   200 & 0.4&0.7&1.5&1.7&0.8&0.9  &  6.9 &  7.5 &  8.5 &  8.4 &  8.3 &  8.7& \multirow{3}{*}{$\frac{p(x)}{g_3(x)}$}\\
   400 &  0.7&0.9&0.0&1.2&1.0&0.9 &  6.9 &  7.9 &  8.6 &  9.8 & 10.1 & 10.1&\\
   800 &  0.6&0.8&0.7&0.6&1.0&1.6 &  7.2 &  8.5 & 10.7 & 13.0 & 14.1 & 16.0&\multirow{2}{*}{$\&$}\\
  1600 &  0.7&0.9&1.0&2.5&2.0&1.2 &  7.4 &  9.2 & 12.4 & 17.1 & 21.3 & 24.7& \multirow{3}{*}{$\frac{3p_0(x)}{g_3(x)}$}\\
  3200 & 0.6&1.2&0.8&2.9&3.2&3.0  &  7.5 & 10.3 & 13.6 & 21.3 & 31.2 & 39.4&\\
  6400 &0.8&1.0&2.0&3.2&2.8&4.6  &  7.5 & 10.3 & 15.2 & 25.7 & 39.4 & 56.7&\\
\hline
\end{tabular}
\end{table}

\noindent
{\it Conclusions to subsection 6.2.2}
\begin{itemize}
\item Tables 7-9\\
In general, the powers of the  new tests  $\hat X^2$ (\ref{stdd3}), $\hat {\check{X}}^2$, $_1\hat {\check{X}}^2$ (\ref{stddc5}) are greater than the power of the test $X^2_{p2}$ (\ref{pois2}) \cite{zex}  developed  for Poisson  histograms.
\item
$X^2_{p2}$ (\ref{pois2}) \cite{zex}  is  a test for comparing  {\it equivalent number of unweighted events}  histograms  and it  cannot  be  directly  interpreted  as a test  for comparison of  original  weighted histograms.     
\end{itemize}

 As a summary, the numerical examples demonstrate superiority of the new tests for comparison of weighted histograms
 under existing tests \cite{gagucom1, gagucom2, zex} including  test applications for Poisson weighted histograms.

\section{Conclusions}
A review of the chi-square homogeneity tests for comparison of weighted  histograms  is presented in this work.
 Bin content of a weighted histogram  is considered as a random sum of random variables that  permit generalization of the classical homogeneity chi-square  test for  histograms with weighted entries. Improvements of  the chi-square tests with better statistical properties are proposed.

  Evaluation of the size and power of  tests is done numerically for different types of weighted histograms with a  different number of events and different weight functions.  In  general, the size of the new tests is  closer to its nominal value  and it is plausible that the power is greater than their  power of currently available tests. The presented numerical examples demonstrate  the superiority of the new  tests over the previously proposed tests for Poisson weighted histograms.

   The proposed tests can be used to fit Monte Carlo data to experimental data, to compare experimental data with Monte Carlo data and to compare two Monte Carlo data sets as well as  to solve the unfolding problem by reweighting the events.

\end{document}